**Engineering Disordered Many-particle Plasmonic Nanoclusters for Wafer-scale Uniform and Giant Electromagnetic Field Enhancement**

Minjun Kim, Vasanthan Devaraj*, Hyeon-Seok Seo, Seongjae Eom, Jeong-Su Lee, Donghan Lee, Min Yong Jeon*, Thomas Zentgraf*, Jong-Min Lee*

Minjun Kim, Min Yong Jeon
Institute of Quantum Systems, Chungnam National University, Daejeon, 34134, Republic of Korea
E-mail: myjeon@cnu.ac.kr

Vasanthan Devaraj, Thomas Zentgraf
Department of Physics & Institute for Photonic Quantum Systems (PhoQS), Paderborn University, Paderborn, 33098, Germany
E-mail: vasanthan.devaraj@uni-paderborn.de
E-mail: thomas.zentgraf@uni-paderborn.de

Hyeon-Seok Seo, Seongjae Eom, Jong-min Lee
School of Nano Convergence Technology & Nano Convergence Technology Center, Hallym University, Chuncheon, 24252, Republic of Korea
E-mail: jmlee@hallym.ac.kr

Jeong-Su Lee, Donghan Lee
Bright Quantum incorporated, Daejeon, 34134, Republic of Korea

Min Yong Jeon
Department of Physics, Chungnam National University, Daejeon, 34134, Republic of Korea

**Abstract**

Scalable plasmonic technologies face a critical trade-off: few-body architectures offer high enhancement but are sensitive to fabrication flaws, while scalable methods like solid-state dewetting yield large, low-enhancement gaps. We introduce a paradigm shift using a many-body plasmonic architecture inspired by statistical mechanics. By moving toward the continuum limit (N>>1), local geometric variations are statistically averaged out, effectively decoupling optical performance from microscopic disorder. We implement this concept via a lithography- and etching-free, multi-step dewetting strategy, creating wafer-scale nanoclusters. This process strategically forms a robust many-body system by introducing numerous small satellite nanoparticles between larger particles. Crucially, this design achieves a high collective enhancement that surpasses even optimized few-body systems, despite having larger individual gaps. Under optimized conditions, these substrates exhibit a surface-enhanced Raman scattering enhancement factor approaching $4 \times 10^8$ with unprecedented reproducibility (RSD of ~10%). This scalable, low-cost concept establishes a practical route toward reproducible wafer-scale nanophotonic platforms for sensing, spectroscopy, and quantum technologies.

M.K, and V.D. contributed equally to this work

## 1. Introduction

Plasmonic nanostructures act as an attractive platform for multitude of applications owing to their remarkable optical[1,2], catalytic[3], and electronic properties[4,5]. Among these, near-field enhancement[6] (NFE) plays an important role in plasmonics by utilizing the strong light-matter interaction, progressing advancements in applications like surface-enhanced spectroscopy[1,6], disease diagnostics[2,7], quantum communication[8], food safety[9], renewable energy[3,4], and environmental monitoring[10]. Among the various strategies to achieve strong NFE with plasmonic nanostructures, precise gap distance control between metallic nanoparticles is highly effective[1,11]. However, translating this nanoscale engineering capability to wafer-sized plasmonic substrates while maintaining uniform NFE remains a major bottleneck, restricting the scalability and commercial adoption of these plasmonic nanostructures.

To date, several approaches have been explored to fabricate surface-enhanced Raman scattering (SERS) substrates, primarily through lithography-based techniques[1,2,8,11] or self-assembly methods[1,6,7,12,13]. Lithography based fabrication offers high precision, excellent reproducibility and stability, but its multiple processing steps lead to higher production costs and long production times. In contrast, self-assembly provides a cost-effective and scalable solution; however, it suffers from key limitations such as non-uniform nanoparticle spacing, poor reproducibility, susceptibility to oxidation, and surface functionalization-induced refractive index variations – factors that collectively diminish NFE uniformity[14,15,16,17,18].

Despite significant advancements in fabrication strategies, there remains a critical gap in the development of wafer-scale plasmonic substrates[19,20,21] that ensure uniform NFE while maintaining mass manufacturing compatibility. A highly efficient large area plasmonic substrate must satisfy the following criteria: (i) strong and uniform NFE over large area; (ii) high reproducibility, and sensitivity; (iii) structural reliability and stability; (iv) a cost-effective and scalable fabrication approach; and (v) compatibility with industrial production standards. The primary challenge lies in achieving plasmonic field enhancement uniformity while maintaining a cost-effective and scalable fabrication process.

To address these challenges, we propose and demonstrate a plasmonic nanogap structure suitable for a self-assembly approach by utilizing the dewetting strategy to attain large area uniformity with minimal processing complexity. The approach reduces equipment requirements, ensures reproducible plasmonic nanostructures formation, consistent optical performance, and these properties were validated through optical, structural, and numerical analyses.

Critically, we challenge a prevailing assumption in the field. Prior studies have emphasized variations in nanoparticle (NP) size and interparticle spacing inevitably degrade optical uniformity, suggesting precise nanoscale control is essential. In contrast, we demonstrate for the first time that wafer-scale uniformity in plasmonic response is instead governed by the statistical consistency of NP size and gap distributions. We show that uniform wafer-scale optical properties can be achieved by preserving statistical consistency in nanoparticle size and gap distributions, even when individual features remain random. This insight resolves a longstanding bottleneck in plasmonics, eliminating the need for costly nanoscale perfection and enabling practical wafer-scale fabrication.

In addition, we demonstrate the ability of our plasmonic nanogap structures to enhance quantum emitter performance, highlighting their potential in quantum photonics. Beyond quantum applications, their high reproducibility, cost efficiency, and wafer-scale uniformity make them promising platforms for biosensing, molecular diagnostics, advanced spectroscopy, and environmental monitoring. Our self-assembly based fabrication strategy overcomes the limitations of previous conventional methods and open insights for cost-effective pathway in making highly efficient plasmonic substrates at wafer scale, thereby expanding its applicability in next-generation sensing, imaging, and quantum technologies.

## 2. Results and Discussion

### 2.1. Addressing the gap size non-uniformity between nanoparticles

We introduce a paradigm shift in the design of plasmonic architectures by leveraging a foundational principle in statistical mechanics governing the transition from few-body to many-body systems. In the few-body regime (N~1), such as a conventional plasmonic dimer, the electromagnetic response is acutely sensitive to microscopic parameters – specifically, the precise interparticle distance or gap size “g”. However, drawing upon the established concept of the continuum limit (or thermodynamic limit)[22], as the number of interacting elements increases (N>>1), microscopic fluctuations are inherently suppressed, and local details are statistically averaged out. In this regime, the collective behavior is dominated by the statistical distribution of interactions rather than the specifics of any individual pairwise coupling. While this principle is fundamental to thermodynamics, applying it to the design of plasmonic nanoclusters represents a significant conceptual shift. We utilize this established principle of statistical averaging to develop plasmonic architectures that are inherently robust against local geometric variations, thereby addressing the critical trade-off between performance and manufacturability. This approach directly confronts limitations in scalable fabrication methods like solid-state dewetting. While enabling wafer-scale production, dewetting typically yields large average interparticle spacings (50 – 100 nm), precluding the intense field confinement associated with idealized sub-10 nm gaps.

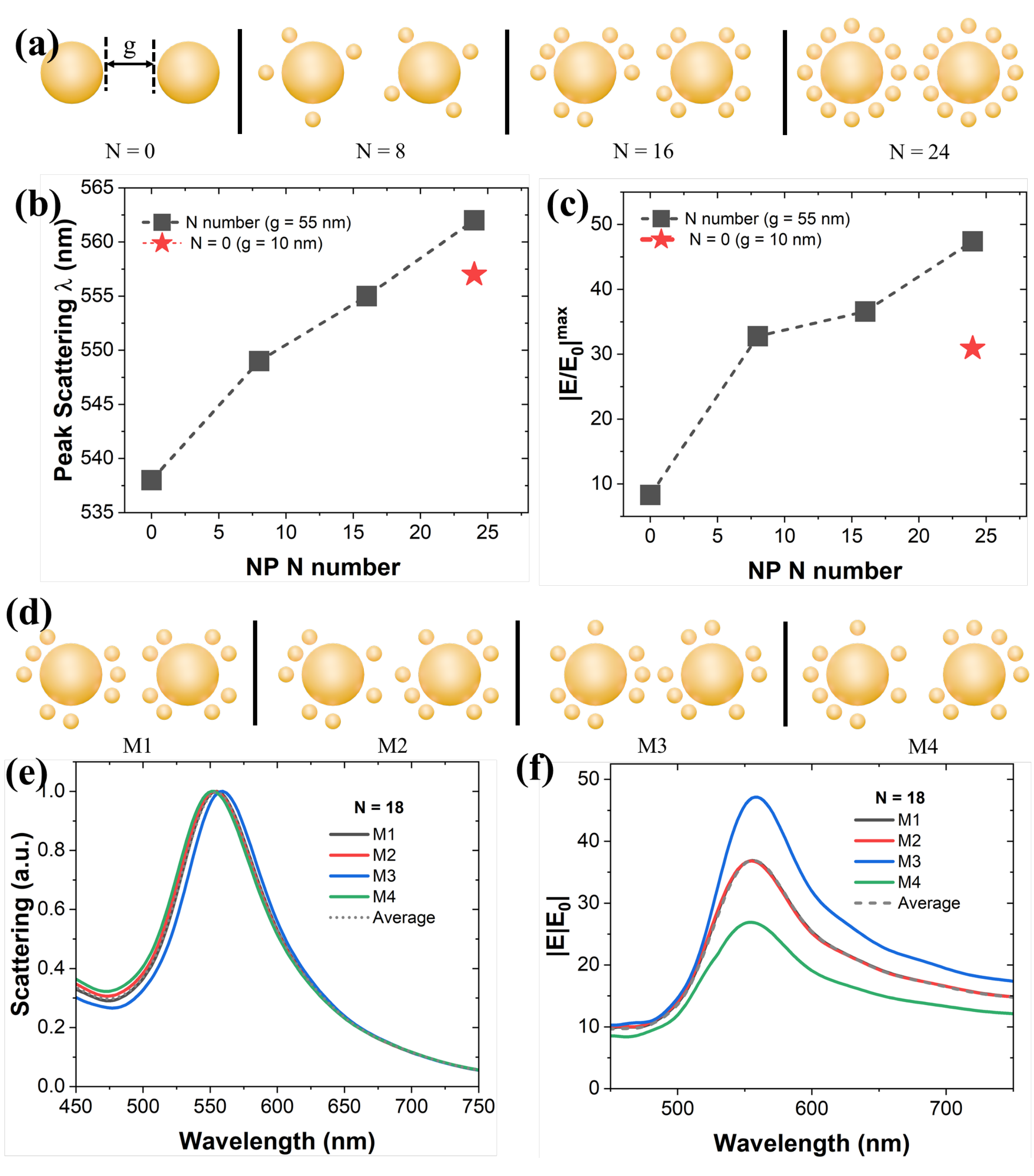

**Figure 1.** (a) Many-body plasmonic cluster architecture with varied N number of satellite

smaller nanoparticles (NPs). Diameter: small NP = 20 nm and big NP = 80 nm. Peak resonance scattering wavelength (b) and maximum near-field enhancement NFE $|E/E_0|$ from (a) models. (d) Varied spatial configuration of an N = 18 cluster and corresponding plasmonic scattering (e) and NFE (f) results. All of these models are simulated at free space environment (n = 1).

To validate this concept, we investigated a satellite nanocluster architecture (Fig. 1a). The system consists of larger (diameter = 80 nm) core NPs separated by a g = 55 nm gap distance and applied a varying number (N = 8, 16, 24) of smaller (diameter = 20 nm) satellite NPs, representing the dewetting constraints. Our analysis reveals a monotonic increase in both the peak scattering wavelength and the near field enhancement (NFE) |E/E0| as N increases (solid black squares, Fig. 1b, c). Remarkably, beyond a certain threshold of N, the collective enhancement generated by the many-body cluster surpasses that of an optimized plasmonic dimer (diameter = 80 nm) with a significantly smaller gap (g = 5 nm) as displayed in Figure 1b,c (solid red star). Detailed spectral responses of figure 1b,c are provided in Supporting Information or SI Figure S1. This demonstrates that the statistical dominance inherent in the many-body approach effectively overcomes the limitations imposed by large individual separations. Crucially, we evaluated the system's robustness against structural disorder, essential for ensuring wafer-scale uniformity. We analyzed the effects of varying the spatial configuration of an N = 18 cluster (Fig. 1d). The results (Fig. 1e & 1f) demonstrate exceptional stability; the overall plasmonic (scattering) response is remarkably insensitive to the specific arrangement of the satellites. Even extreme local variations that induced a twofold difference in local NFE did not result in a discernible spectral shift. Two significant observations can be noted here considering wafer-scalable fabrication in terms of NFE: (i) With N = 18 model, averaging (dotted grey line – Fig. 1e,f) of plasmonic responses becomes dominant due to the uneven distribution of satellite NPs filling; (ii) When there are sufficient filling of smaller NPs (example N= 24), spatial uniformity in NP distributions will be improved and can critically addresses the main concern, highly uniform NFE.

These findings confirm that the many-body plasmonic cluster design, by relying on established statistical principles, successfully decouples optical performance from local disorder, simultaneously enabling high enhancement factors and exceptional uniformity. Our approach demonstrates the critical importance of small (satellite) NPs when introduced to a plasmonic substrate with non-uniformly distributed dimer NPs. Our concept of introducing satellite NPs in such plasmonic substrates will address the following concerns: (i) Uniformity in LSPR resonances; (ii) Uniformity in NFE signal; and (iii) possibly achievable with simple fabrication approach (no need of lithography support). Significantly, all the above merits are attainable over a large area of a substrate, even up to wafer scale fabrication. This means, highly uniform NFE and LSPR signals attainable on a large substrate will benefit wide variety of on-demand applications.

**2.2. Fabrication concept and experimental approach to realize satellite-structured gold nanoparticles**

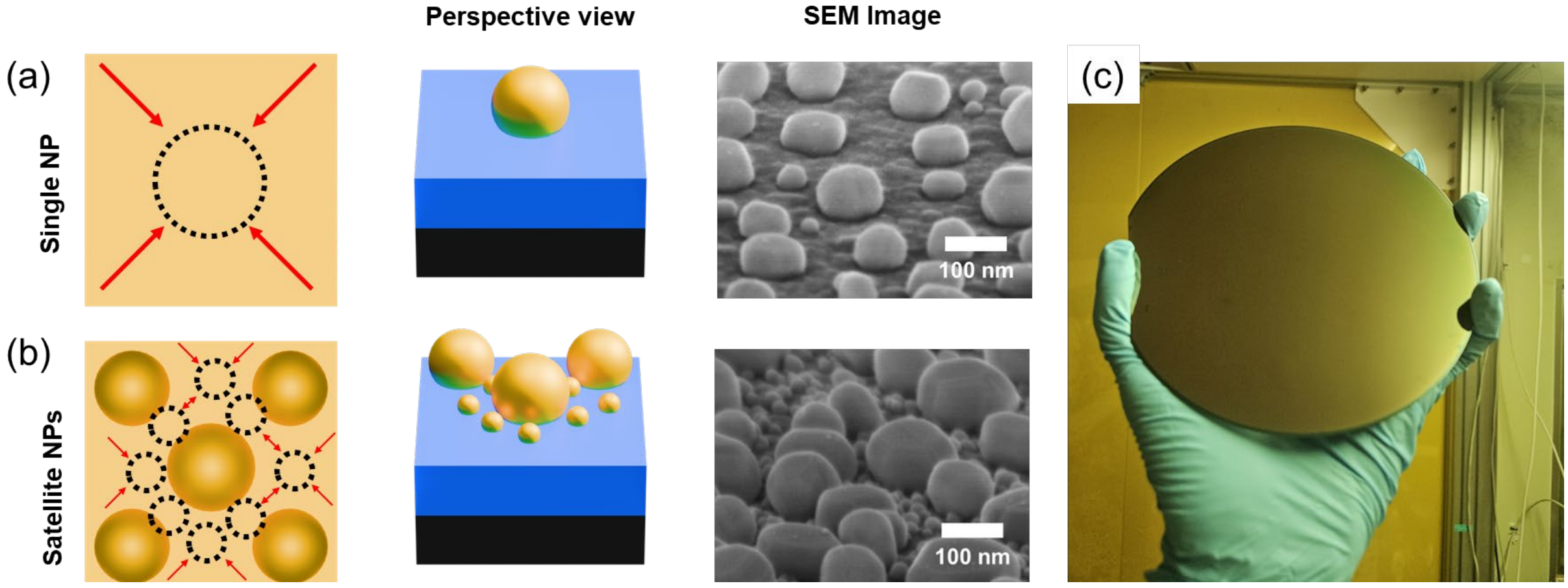

**Figure 2**. Schematic illustrations and corresponding scanning electron microscopy (SEM) images detailing the realization of satellite-structured gold nanoparticles (Au NPs). (a) Formation of primary Au NPs. A thin gold film undergoes solid-state dewetting, leading to the self-assembly of isolated NPs with a random size distribution on the substrate. The red arrows indicate the direction of material retraction. (b) Formation of satellite-structured Au NPs via a sequential process. A subsequent gold film is deposited over the pre-existing primary NPs and subjected to a second dewetting step. The dewetting of this second layer is spatially constrained by the limited available substrate area and the elongated configuration of the gold film surrounding the primary structures. This directed assembly results in the preferential formation of smaller, secondary (satellite) Au NPs localized immediately adjacent to the primary NPs. (c) Optical image of a 6-inch Si wafer with gold nanoparticles produced through the triple-dewetting process.

Large-area self-assembly processes offer a compelling route to commercialization by substantially reducing the processing cost per unit device, a critical advantage over conventional nanopatterning lithography. A promising lithography-free approach involves fabricating metal nanoparticle (NP)-based satellite structures by precisely controlling material supply during a sequential, or 'dual', solid-state dewetting process (Fig. 2, Supporting

Information or SI Fig. S2). During the formation of metal NPs through dewetting, the particles attract metal from areas larger than their diameter to achieve sufficient volume. In the first dewetting process, metal NPs form at regular intervals. When additional (second) deposition and dewetting are performed on a substrate that already has NPs, the existing NPs slightly increase in size, while new NPs form in the vacant spaces. If the vacant spaces are sufficiently wide, NPs similar in size to those formed during single dewetting with the same metal film thickness can develop. Conversely, if the vacant spaces are narrow and elongated, the amount of metal available for aggregation decreases, resulting in the formation of numerous smaller NPs. By appropriately controlling the NP density through dual dewetting, interparticle gap areas can be created where smaller NPs can form.

### 2.3. SERS Enhancement of Satellite Nanostructures using Triple-Dewetting

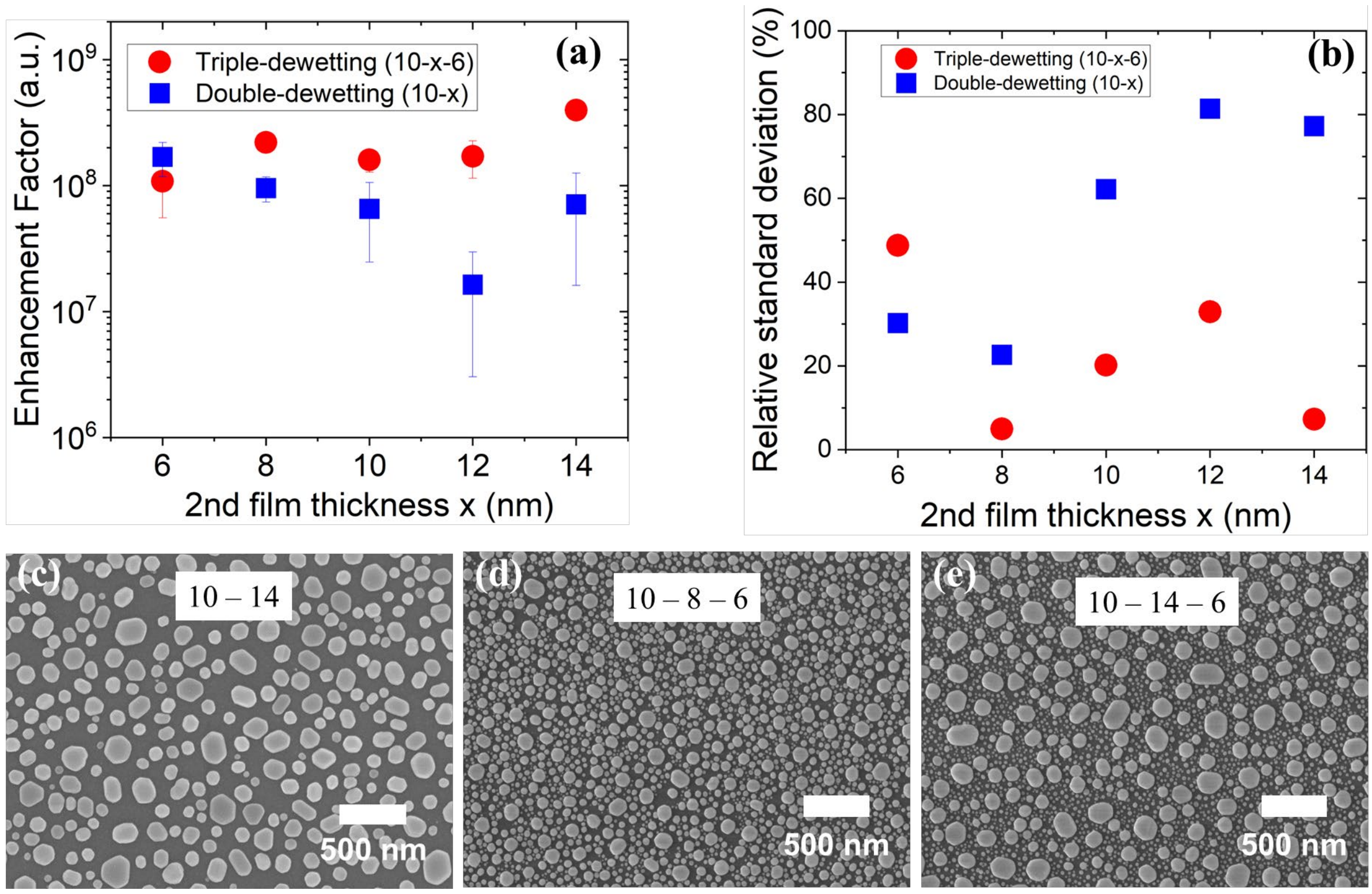

**Figure 3.** (a) Experimental SERS enhancement factors and (b) relative standard deviation percentage of SERS enhancement as measured from plasmonic nanostructures fabricated via double and triple-dewetting strategy (naming parameter: ex. 10-14-6: first film thickness condition is 10 nm, second film thickness condition is 14 nm, and third film thickness condition is 6 nm). The best condition for uniform near-field enhancement is achieved with the 10-14-6 triple de-wetting approach. SEM images of dewetted plasmonic nanostructures from following conditions: (c) 10-14, (d) 10-8-6, and (e) 10-14-6.

A dewetting process optimization was performed to implement the process design strategy in Figure 2. For the first dewetting, a 10 nm thick gold film was used, which is known to uniformly form nanoparticles (NPs) with an average diameter of approximately 75 nm, suitable for the visible light range. For the third dewetting, a 6 nm thick gold film was employed, a condition known to produce a large number of smaller NPs to facilitate the formation of satellite structures (SI Fig. S2). The gold film thickness for the second dewetting was carefully controlled. The maximum thickness was limited to 14 nm because NP formation becomes unstable beyond this point. Raw SERS measurements measured from these double and triple-dewetting conditions are provided in SI Figures S3 and S4. From these raw SERS data, we evaluated the SERS enhancement factor (Fig. 3a) and the corresponding relative standard deviation of the SERS signal (Fig. 3b). When comparing double dewetting and triple dewetting, it is critical to consider both the SERS enhancement factor and its uniformity across the substrate. We used following naming notations for the dewetting conditions throughout the manuscript: (i) double dewetting (first film thickness – second film thickness: example, 10-6), and (ii) triple dewetting (first film thickness – second film thickness – third film thickness: example, 10-6-6). All the film thickness conditions are in nanometers. Although the 10-6 condition (double dewetting) yields high enhancement factors, triple dewetting achieves both high SERS enhancement factors and excellent spatial uniformity. This uniformity is quantitatively reflected by the relative standard deviation (RSD), an essential metric for wafer-scale fabrication. It directly indicates the reproducibility and reliability of the plasmonic substrate across large areas – factors compulsory for practical and commercial adoption. Specifically, for double dewetting, the initially large gaps between nanoparticles (NPs) formed during the first step result in a broad and inconsistent size distribution in the subsequent formation of smaller NPs, severely deteriorating uniformity. In sharp contrast, triple dewetting first produces densely packed, larger NPs, subsequently facilitating the controlled placement of smaller satellite NPs into the interparticle gaps. This satellite NP approach significantly reduces gap-size variability, resulting in giant, uniform SERS signals across extensive substrate areas, thereby fulfilling the crucial criteria for scalable plasmonic device manufacturing. We selected three types of dewetting: 10-14, 10-8-6, and 10-14-6 to examine the characteristics of self-assembled nanoparticle clusters produced by different dewetting methods (complete set of SEM data available in SI figures S5 – S7). Figure 3 (c – e) shows SEM images of 10-14, 10-8-6, and 10-14-6 respectively, illustrating the nanoparticle distributions formed via different dewetting processes. The 10-14-6 system exhibits a high density of satellite nanoparticles surrounding a larger central particle, as evidenced in SI Figure S8 (a), where the particle size distribution within a 7.5 $\mu m^2$ area follows a Gaussian distribution in the 50-150 nm range, with thousands of smaller satellites below 50 nm. The presence of these densely packed smaller nanoparticles with uniformity spread across the substrate results in smaller interparticle distances (SI Fig. S7e). In contrast, the 10-8-6 system, consisting of larger primary particles, exhibits a broader size distribution with fewer satellite formations (SI Fig. S8b). This structural configuration results in reduced frequency of nanoscale gaps, as reflected in SI figure S7, potentially affecting SERS uniformity. The 10-14 system, which lacks the final dewetting step, demonstrates the absence of satellite formation (SI Fig. S8c) and exhibits large interparticle distances, limiting plasmonic enhancement. These geometrical distribution analyses are directly aligning with our experimental relative standard deviation results.

### 2.4. Understanding why triple-dewetting conditions fare well in uniform near-field enhancement

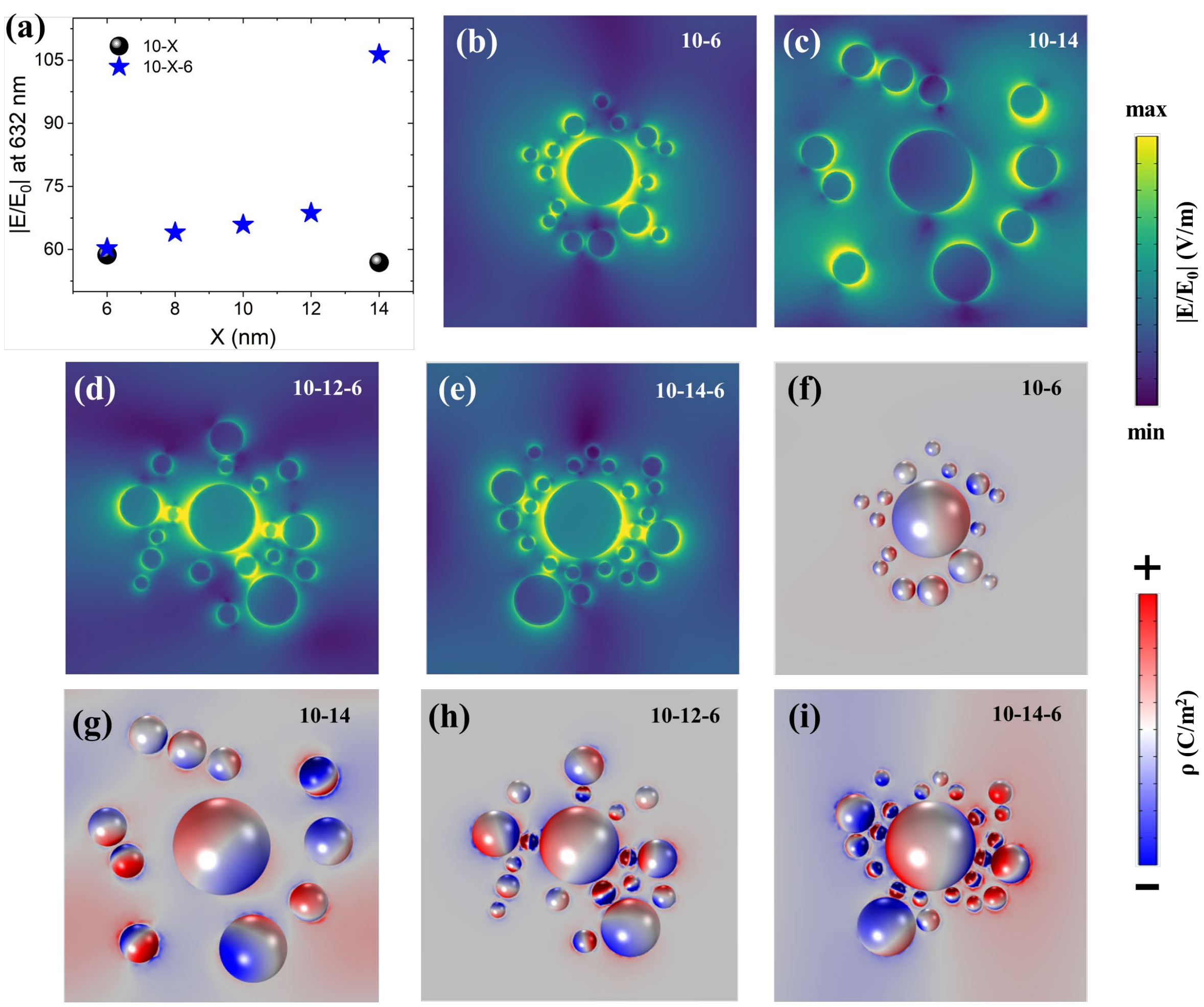

**Figure 4.** Simulation analysis of double and triple dewetting plasmonic nanostructure models. (a) Near-field enhancement $|E/E_0|$ extracted at 632 nm from double and triple dewetting models. Cross-sectional (XY) electric field amplitude profiles (b – e) and corresponding three-

dimensional surface charge density distribution mappings (f – i) as extracted from double and triple dewetting models at 632 nm. The color scale strength is kept same for electric field intensity profiles and surface charge density mappings for a fair comparison.

To further understand the optical results as shown with figure 3(a, b), we simulated plasmonic models by considering the NP distributions as observed from the SEM images of double and triple dewetting conditions (SI Fig. S6 and S7). Simplified model are created based on how the NPs are distributed according the dewetting conditions (SI Fig. 9). Figure 4a shows the simulated near field enhancement $|E/E_0|$ at 632 nm from double and triple dewetting NP models on a $SiO_2$/Si substrate. These data points are extracted from the near-field spectra as shown in SI figure S10 (a, b). In case of double dewetting models, the near field enhancement factor is ~ similar for 10-6 and 10-14. The near field enhancement factor started to increase slowly from 10-6-6 to 10-12-6 and a rapid increase at 10-14-6 is observed. The simulation trend for the NFE for double and triple dewetting models agrees well with the experimental results. To understand further, why larger near field enhancement happens with 10-14-6 than the rest of the NP distribution models, we plotted cross-sectional XY (XY cut at NP-$SiO_2$ interface) electric field amplitude profiles (Fig. 4 b – e) and the corresponding three-dimensional surface charge density mappings (fig. 4 f – i). For a 10-14 model, even though there are few NPs close to each other, the non-uniform distribution of NPs results in poor near-field enhancement distribution over a large area. This is the reason why 10-14 exhibits a relative large standard deviation of SERS experimentally.

An interesting comparison comes between 10-6 and 10-14-6 model (SI Fig. S10 c – d). Both of these plasmonic substrates feature comparable distributions of satellite (small) NPs. The major difference lies in the distribution and distance separation between the big NPs. With 10-14-6, we have relatively closer spacing between big NPs as compared with 10-6. To demonstrate the differences, we simulated a free-space satellite model (single versus dimer NP, SI Fig. S10 e – g). The NFE for the dimer satellite model displayed ~ 1.6 times increment in near-field enhancement than the single NP model. Electric field amplitude profiles extracted at the resonance peak wavelength positions showed better hot-spot localized field enhancement with dimer satellite model (SI Fig. S10g). With 10-12-6 as compared with 10-14-6, the distribution between large NPs to smaller NPs is still not comparably uniform.

Three-dimensional surface charge density maps further strengthen our interpretation (Fig. 4 f – i). With 10-14 model, even though, we have a dipolar mode for a single or dimer or few NPs close to each other, the overall charge distribution spread on the substrate becomes non-dipolar. With the 10-6 model, weaker dipolar characteristics are noted as compared with 10-14-6. With 10-14-6, better dipolar mode characteristics originates on the basis of big NPs – satellite NPs arrangement displaying better uniformity in gap-size distribution and closer big NP spacing. In other words, the best triple dewetting condition yields high and uniform SERS enhancement with lesser relative standard deviation over a large area of the substrate.

### 2.5. Demonstration of large area SERS uniformity from satellite plasmonic structures with Raman mapping

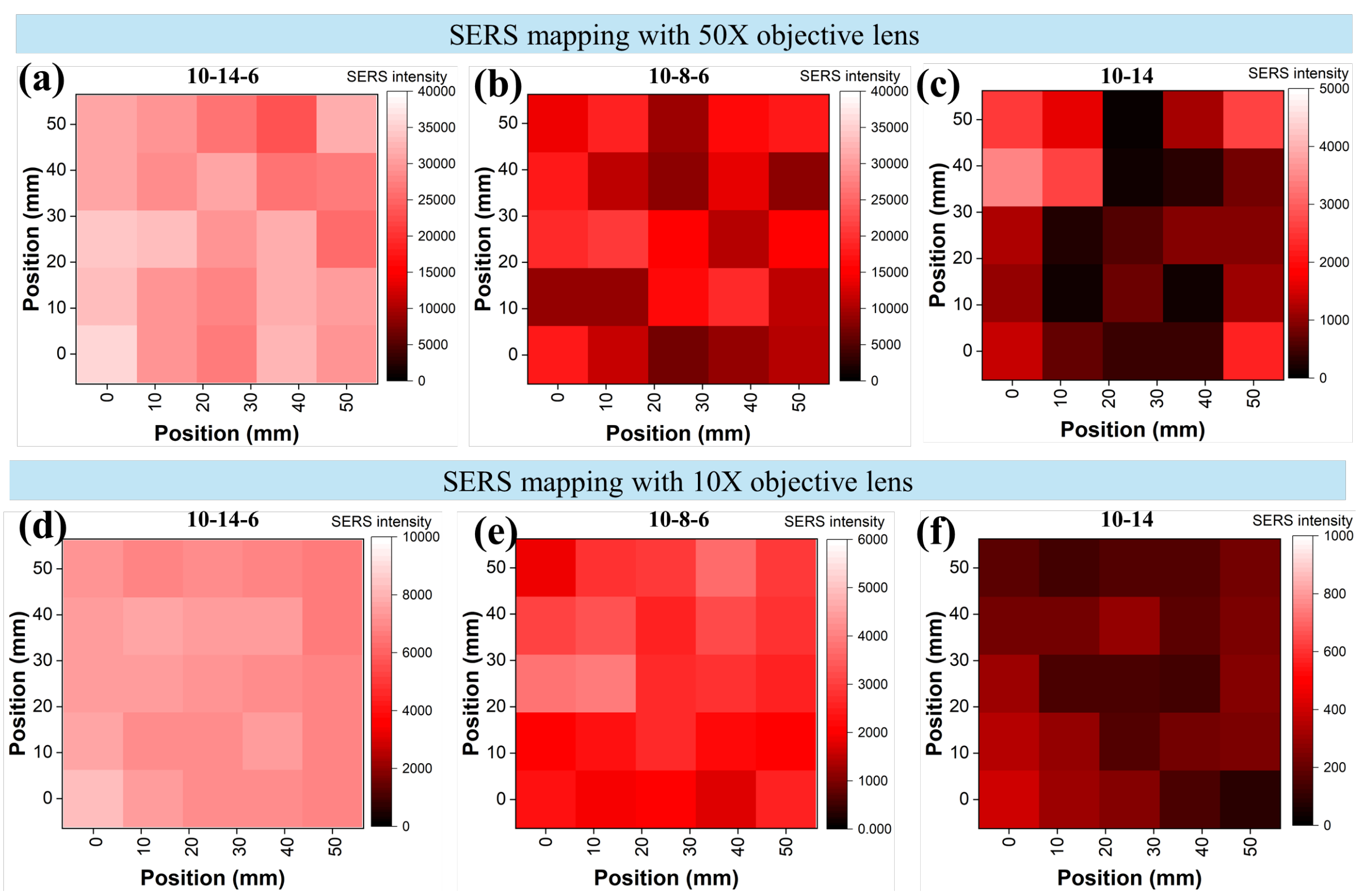

**Figure 5.** Mapping results of benzenethiol 1072 cm$^{-1}$ of large-area samples (a) 10-14-6, (b) 10-8-6, (c) 10-14, as measured with a 50X objective lens excited by a 633 nm laser, and (d) 10-14-6, (e) 10-8-6, (c) 10-14, as measured with a 10X objective lens. SERS intensity unit is counts/s.

To evaluate the applicability of integrated devices, large-area Raman mapping was conducted using both laser spot sizes of 0.8 μm$^2$ (50X objective lens) and 7.5 μm$^2$ (10X). Mapping was performed using benzenethiol's 1072 cm$^{-1}$ Raman peak over a 5 × 5 cm$^2$ area under 633 nm laser excitation (Fig. 5 a – f). For complete SERS spectra of these plasmonic substrates see, SI Figure 11. In self-assembled plasmonic structures, uniformity is typically evaluated over a wide laser spot size. However, ensuring excellent uniformity at small laser spot sizes is crucial for integration with nanodevices. With 0.8 μm$^2$ laser spot size, the triple-dewetting 10-14-6 substrate exhibited the highest SERS enhancement factor (EF = $3.9 \times 10^8$) coupled with an outstanding uniformity, reflected by a low RSD of 9.63% (Fig. 5(a)). In contrast, the 10-8-6 substrate, characterized by fewer satellite nanoparticles, showed a moderate EF of $1.5 \times 10^8$ and a higher RSD of 22.2% (Fig. 5b). The double-dewetting 10-14 substrate, having the largest interparticle spacing, exhibited a significantly lower EF of $1.5 \times 10^7$ and a very high RSD of 83.5% (Fig. 5c). With 7.5 $\mu m^2$ laser spot size, the 10-14-6 substrate maintained excellent uniformity with an even lower RSD of 7.28% (Fig. 5d), while the 10-8-6 and 10-14 substrates exhibited notable improvements in uniformity (RSD decreased to 12.7% and 39.7%, respectively) (Fig. 5 e – f). This indicates that densely populated satellite structures, as in the 10-14-6 system, effectively ensure consistent SERS signals across varying spatial scales. Typically, random or sparse nanostructures display reduced RSD at lower magnifications due to increased sampling area, whereas the highly uniform 10-14-6 structure shows only marginal improvement, signifying inherently superior uniformity at all measured scales. Notably, the 10-14-6 sample demonstrated exceptional signal homogeneity, maintaining a relative standard deviation (RSD) below 10% even when analyzed with the smaller 0.8 $\mu m^2$ spot size. Given the difference in the laser spot size, this high degree of uniformity suggests that the 10-14-6 substrate potentially allows for up to a 10-fold increase in production throughput per wafer compared to the other samples.

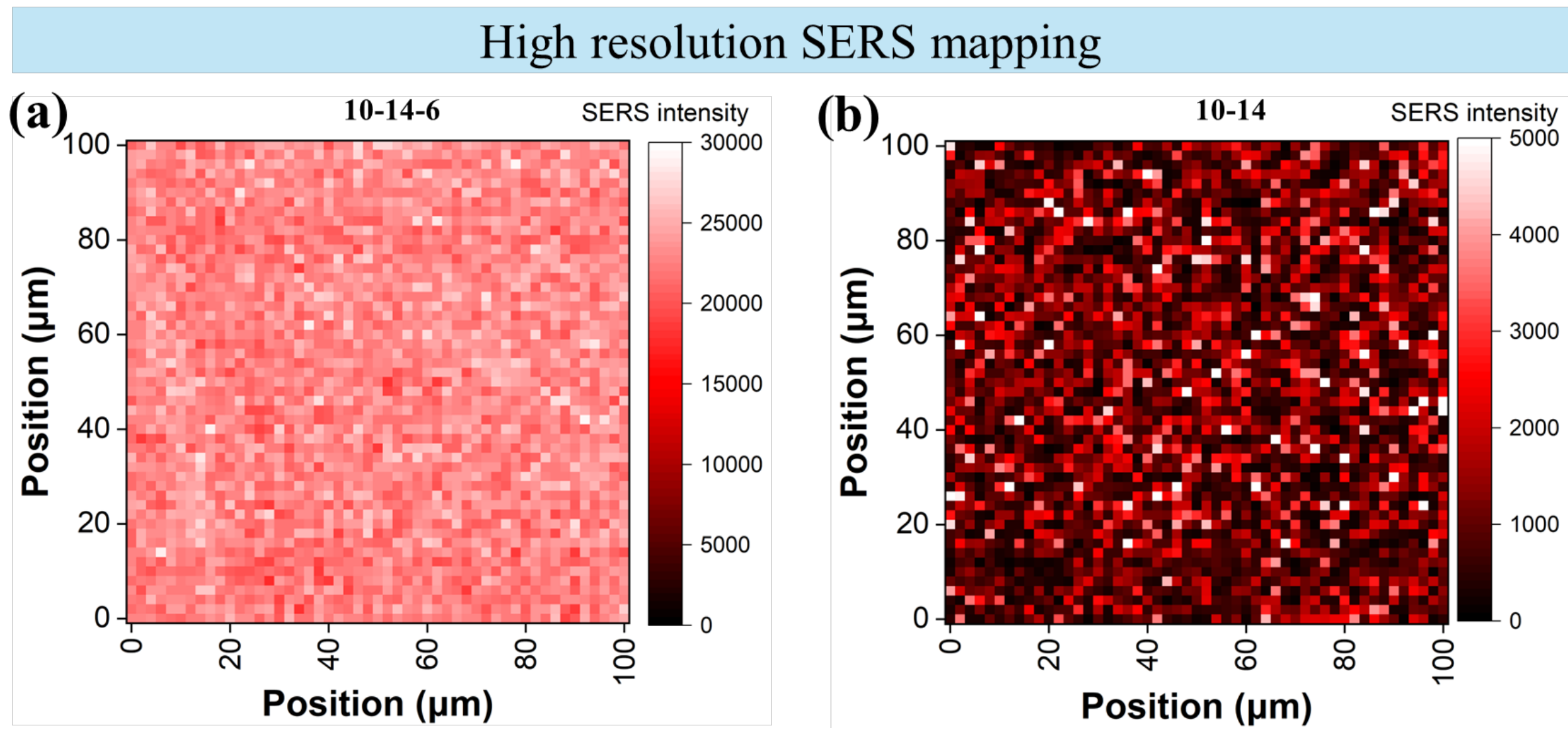

Figure 6. High-resolution SERS mapping of benzenethiol (1072 $cm^{-1}$) measured at 2 μm intervals for (a) 10-14-6 and (b) 10-14 samples. SERS intensity unit is counts/s.

To further validate the uniformity and hotspot distribution, high-resolution Raman mapping was conducted at 2 μm intervals, closely matching the laser spot size (complete SERS spectra are provided in SI figures S12 and S13). The optimized 10-14-6 sample exhibited a remarkably low RSD of 7.57%, clearly demonstrating exceptional local uniformity (Fig. 6a). In a stark contrast, the 10-14 sample showed a significantly higher RSD of 93.13%, indicating severe spatial heterogeneity and inconsistent hotspot distribution (Fig. 6b). We also carried out statistical analyses collected from 30 SEM images for 10-14, 10-8-6 and 10-14-6 dewetted plasmonic nanostructures corresponding to 7.5 $\mu m^2$ (10X) and 0.8 $\mu m^2$ (50X) mapping area (SI Fig. S14). 10-14-6 displayed better consistency in terms of relative standard deviation (RSD) of average particle size and average interparticle distance across different analysis areas. More importantly, the high-resolution mapping (0.8 $\mu m^2$) of 10-14-6 dewetted nanostructures reveals exceptional local uniformity (SI Fig. S14d), which is an essential parameter for wafer-scalable fabrication. These results provide robust evidence that the densely packed satellite nanoparticle arrangement consistently generates uniform plasmonic hotspots, not only on large scales but also within micro-scale regions. Achieving uniform hotspot distributions at these localized scales is critical for reliable performance for various optical devices and facilitates high-density integration of plasmonic systems.

### 2.6. Enhanced Photoluminescence and Radiative Recombination of Nano-emitter

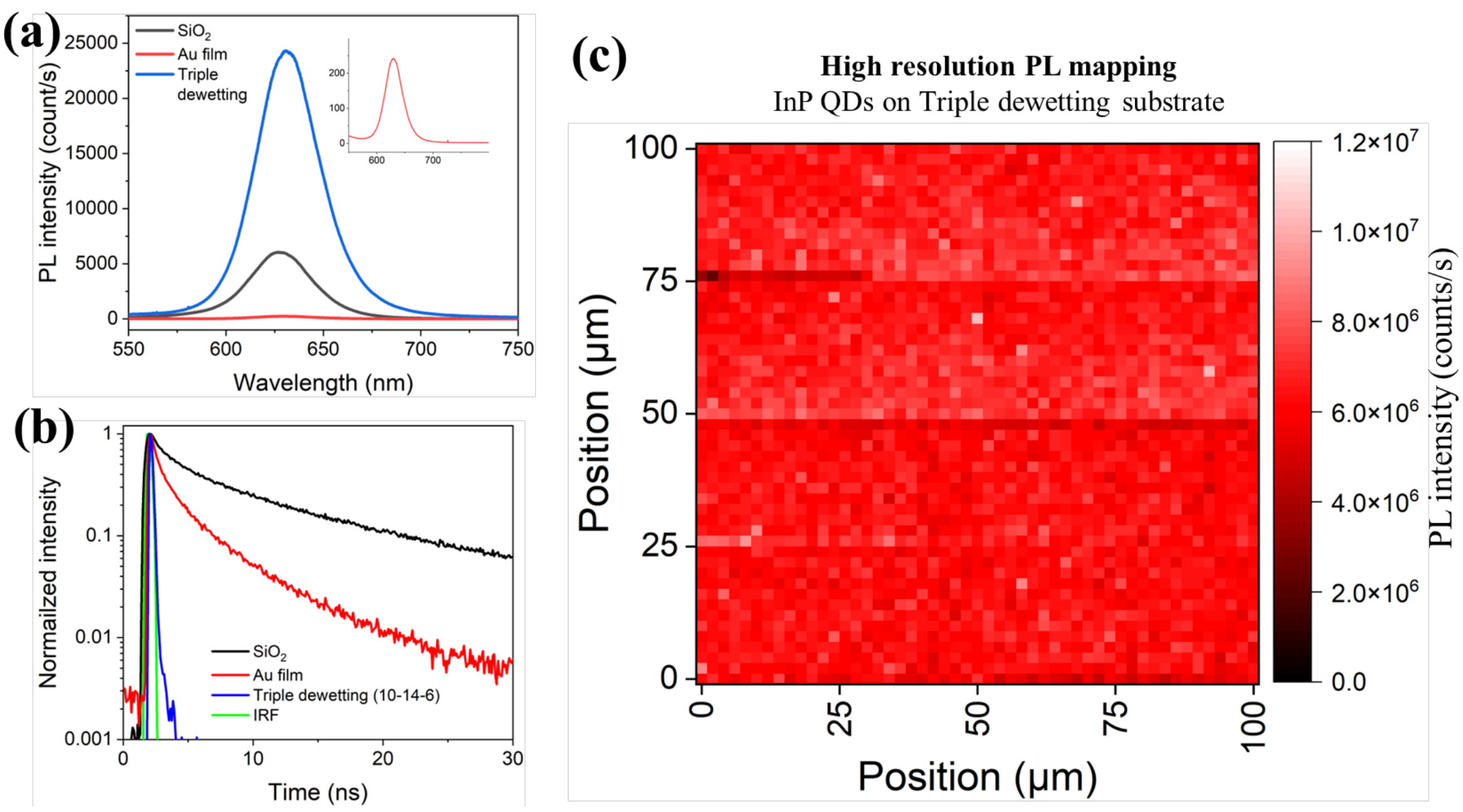

**Figure 7.** Photoluminescence (PL) characterization of InP quantum dots on different substrates. (a) Averaged PL spectra from InP QDs on triple dewetted (10-14-6), $SiO_2$, and Au-coated substrates (inset: magnified InP PL on Au substrate). (b) TRPL decay curves for InP QDs on the three substrates. (c) PL intensity map of InP QDs on a 100 x 100 $\mu m^2$ area of the triple dewetted plasmonic substrate.

The strong electromagnetic field enhancement and its highly uniform distribution, as previously discussed for the triple dewetted (10-14-6) plasmonic substrate platform, provide an ideal environment for developing effective and efficient photonic devices. To probe the potential of this platform, we uniformly introduce InP quantum dots (QDs) onto the sample surface via spin coating. Assuming a uniform coating of QDs, the distribution of plasmonic enhancement can be directly mapped. For comparison, a Si substrate coated with 300 nm of $SiO_2$ and a substrate coated with 100 nm of gold were used as control samples. Figure 7a displays the photoluminescence (PL) spectra measured from each substrate. Each spectrum represents the average of 25 measurements taken at random positions on the substrate under identical conditions (SI Fig. S15). The PL intensity from the triple dewetting substrate (10-14-6) is ~ four times greater than that from the $SiO_2$ substrate and over 100 times greater than that from the gold-coated substrate. Figure 7b shows the results from time-resolved photoluminescence (TRPL) measurements for each substrate. The average PL decay time, calculated by considering the relative amplitudes of the decay components, was determined to be 0.15 ns for the triple dewetting substrate, 20.4 ns for the $SiO_2$ substrate, and 4.16 ns for the gold-coated substrate. The reduction in PL decay time on the gold-coated substrate, compared to the $SiO_2$ substrate, is attributed to quenching. This effect arises from the loss of charge carriers within the QDs through non-radiative recombination pathways, leading to the significant decrease in PL intensity observed in Figure 7a.  In contrast, while the triple dewetting substrate exhibits an even more pronounced reduction in PL decay time, its PL intensity is substantially enhanced. This provides strong evidence for the enhancement of the radiative recombination rate due to plasmonic coupling. The plasmonic structure accelerates the emitter's spontaneous emission by increasing the local density of optical states (the Purcell effect), which enhances the radiative decay rate more significantly than it introduces non-radiative loss channels, thereby increasing the overall photon output. To assess uniformity, PL mapping was performed over a 100 μm × 100 μm area on the triple dewetting substrate with a step size of 2 μm (Figure 7c, complete PL spectra in SI Fig. S16). The relative standard deviation of the integrated PL intensity was found to be 10.8% which again proves an exceptional localized optical uniformity that can be obtained from the triple dewetting substrate.

**3. Conclusion**

This work demonstrates a significant advancement in the fabrication of wafer-scale plasmonic nanoparticle clusters substrates by achieving exceptional optical uniformity through a highly efficient and scalable self-assembly-based process. This fabrication method leverages on controlled spatial variations in dewetting behavior of metal thin-film(s) formation, which governs the strategic positioning of satellite nanoparticles around primary nanoparticles, eliminating the need for complex lithography or etching processes. This simplicity reduces costs, enables high-throughput production, and establishes a commercially viable route toward large-area plasmonic substrates.

Despite the inherent randomness in nanoparticle size and spatial arrangement, statistical analyses reveal highly uniform average characteristics, ensuring consistent optical responses across the wafer. Experimentally, benzenethiol (BT) molecule-coated substrates demonstrated a surface-enhanced Raman scattering (SERS) enhancement factor exceeding $10^8$ with spatial deviation below 8%. In addition, quantum dots integration reveals localized photoluminescence (PL) intensity uniformity within 11% on the triple dewetting plasmonic substrate. Simulations based on particle distribution models corroborated the near-field enhancement trends observed in experimental SERS measurements. Collectively, these findings highlight that wafer-scale uniformity in plasmonic performance is governed by the statistical consistency of nanoparticle size and gap distributions, rather than by the precise nanoscale positioning of individual particles.

Altogether, structural, optical, and numerical analyses provide strong and consistent evidence for the reliability, uniformity, and performance of our plasmonic nanogap structures. The demonstrated degree of uniformity and enhancement underscores the robustness of the fabrication technique and validates self-assembly-based strategies as powerful approaches for developing advanced plasmonic substrates. The robust wafer-scale reproducibility demonstrated here paves the way for practical applications in biosensing, spectroscopy, quantum emitter coupling, and nanophotonics, marking an important step toward the commercialization of next-generation optical technologies.

## 4. Experimental Section

*Simulation*

We employed a 3D electromagnetic Maxwell solver to carry out numerical simulations using ANSYS OPTICS Lumerical FDTD package12. A plane-wave source is utilized to illuminate the plasmonic nanostructures from the top. A perfectly matched layer (PML) boundary conditions are applied in XYZ directions. Scattering spectra is extracted from a box shaped power monitor which is positioned close to the plasmonic nanostructures. Electric-field amplitude profiles are evaluated by positioning cross-sectional XZ and XY field monitors, respectively. A general meshing size of 20 nm is applied in simulated region and a mesh override of 0.5 nm is applied close to metallic nanoparticles. The refractive indices utilized in our simulations are as follows: in air environment (n = 1). Fo Palik (Si and SiO2)[23], Johnson and Christy (Au)[24]. All of our plasmonic structures are modelled r three-dimensional surface charge density distribution (ρ) mappings, we utilized COMSOL Multiphysics solver (Wave optics module)[12]. By considering an outward normal vector (n), skin effect (δ), local electric field (E), and permittivity of the vacuum (ε0), we extracted the ρ:

$$\rho = \frac{\varepsilon_0 \left(n_x . E_x + n_y . E_y + n_z . E_z\right)}{\delta\left(1 - e^{-\frac{R}{\delta}}\right)} \ \alpha \left(n_x . E_x + n_y . E_y + n_z . E_z\right)$$

As shown from above equation, surface charge density ρ is given as ($n_x \cdot E_x + n_y \cdot E_y + n_z \cdot E_z$).

Fabrication of satellite-structured gold nanoparticles

A silicon wafer substrate was initially coated with a 300 nm thick $SiO_2$ layer deposited using plasma-enhanced chemical vapor deposition (PECVD). Subsequently, an electron-beam evaporator (KVE-E2000L, Korea Vacuum Tech, Republic of Korea) was employed to deposit a gold thin film onto the $SiO_2$-coated substrate for the initial dewetting step. The gold-coated substrate was rapidly annealed using a rapid thermal annealing (RTA) system (RTP 5000, Sntek, Republic of Korea), ramping from room temperature to 820°C within 90 seconds, and subsequently maintained at this temperature for an additional 120 seconds to facilitate nanoparticle formation. This deposition-annealing cycle was repeated two more times under identical thermal conditions, adjusting only the gold film thickness for each deposition, to achieve controlled satellite nanoparticle formation.

*SEM and statistical analysis*

The morphology and distribution of fabricated gold nanoparticles were examined using field-emission scanning electron microscopy (S-4800, Hitachi, Japan). To precisely quantify nanoparticle size distributions and interparticle gaps, SEM images captured at 30,000× magnification (2560 × 1920 pixels) were analyzed. Overlapping particles in the SEM images were systematically identified and separated using deep learning-based image segmentation. The You Only Look Once (YOLO) model was used for object recognition, with the version YOLO11m and parameter M = 20.1. Nanoparticle spacing was accurately determined by calculating linear distances based on particle centroid positions and edge coordinates.

Raman analysis

For SERS reference measurements, substrates were immersed for two days in a dilute benzenethiol solution prepared at a concentration ratio of 10 μL benzenethiol to 50 mL ethanol.

Each substrate was rinsed thoroughly with pure ethanol for 3 minutes to form a uniform self-assembled monolayer (SAM). Raman spectroscopy was conducted using a spectrometer (LabRAM HR-800, Horiba France SAS, France) equipped with a 633 nm excitation laser set to a power of 752 μW. The excitation beam was precisely focused and scattered signals were efficiently collected using a 50X working distance objective lens with a numerical aperture of 0.75. Spectral dispersion was achieved using a diffraction grating with 600 grooves/mm.

*Photoluminescence and Time-resolved photoluminescence*

Colloidal indium phosphide (InP) quantum dots (InP-625-100, Mesolight, China), dispersed in octane with a peak emission wavelength of 625 nm, were used as the emitters. Quantum dot dispersions were prepared at a concentration of 25 μg/mL, and 5 mL of solution was deposited onto the substrate by spin-coating. The spin-coating process was carried out at 3,500 rpm for 40 s to ensure uniform coverage across the sample surface.

Photoluminescence (PL) characterization was performed using a microscope spectrometer (LabRam Soleil, HORIBA France SAS, France). Excitation was provided by a continuous-wave 532 nm laser, with a grating of 600 grooves/mm employed for spectral dispersion. The excitation laser power was set to 0.46 mW with an integration time of 1 s for each acquisition. A long-working-distance 50X objective lens with a numerical aperture (NA) of 0.6 was used both to focus the excitation beam onto the sample and to collect the emitted PL signal.

Time-resolved PL (TRPL) and imaging studies were performed using an inverted-type scanning confocal microscope (SP8 FALCON, Leica Microsystems, Germany) at the Korea Basic Science Institute (KBSI), Daegu. A pulsed laser diode (470 nm, 2.5-5 MHz repetition rate; PicoQuant, Germany) was used as the excitation source. Emission from the InP QDs was collected over the spectral range of 570-780 nm using a hybrid photon detector. The PL decay curves were analyzed by exponential fitting with the Leica LAS X software package (version 3.5.2), yielding average lifetimes for each sample.

**Acknowledgements**

M.K, and V.D contributed equally to this work. This research was supported by the Basic Science Research Program through the National Research Foundation of Korea (NRF) funded by the Ministry of Education (RS-2020-NR049597), Global - Learning & Academic research institution for Master's·PhD students, and Postdocs(G-LAMP) Program of the National Research Foundation of Korea(NRF) grant funded by the Ministry of Education(No. RS-2025-25442707), the funding from the National Research Foundation of Korea (NRF) grant (No. RS-2023-00219703), and the Korea Environmental Industry & Technology Institute (KEITI) grant (RS-2022-KE002262). V.D. and T.Z. acknowledge financial support by the German Ministry of Education and Research (BMBF) within the PhoQuant project (grant number 13N16103).

**Supporting Information**

Supporting Information is available from the Wiley Online Library.

This work presents a lithography- and etching-free dewetting strategy that assembles many-particle gold nanostructures into uniform plasmonic substrates at wafer scale. By forming satellite nanoparticles between larger ones, the method achieves giant and highly reproducible electromagnetic enhancement. The approach establishes a cost-effective route for practical SERS sensing and quantum photonic integration.

**Engineering Disordered Many-particle Plasmonic Nanoclusters for Wafer-scale Uniform and Giant Electromagnetic Field Enhancement**

ToC figure

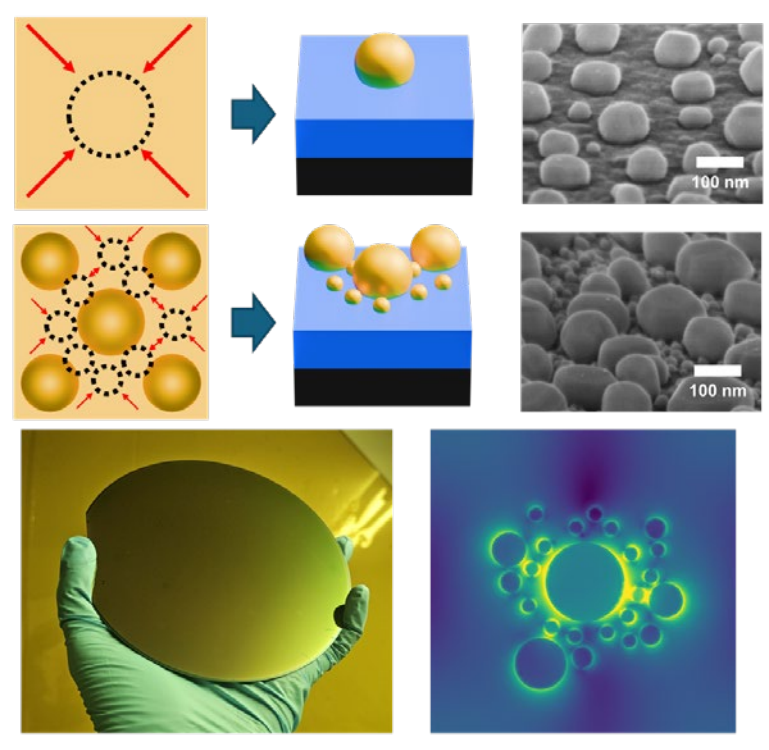

# Supporting Information

**Engineering Disordered Many-particle Plasmonic Nanoclusters for Wafer-scale Uniform and Giant Electromagnetic Field Enhancement**

Minjun Kim, Vasanthan Devaraj*, Hyeon-Seok Seo, Seongjae Eom, Jeong-Su Lee, Donghan Lee, Min Yong Jeon*, Thomas Zentgraf*, Jong-Min Lee*

Minjun Kim, Min Yong Jeon
Institute of Quantum Systems, Chungnam National University, Daejeon, 34134, Republic of Korea
E-mail: myjeon@cnu.ac.kr

Vasanthan Devaraj, Thomas Zentgraf
Department of Physics & Institute for Photonic Quantum Systems (PhoQS), Paderborn University, Paderborn, 33098, Germany
E-mail: vasanthan.devaraj@uni-paderborn.de
E-mail: thomas.zentgraf@uni-paderborn.de

Hyeon-Seok Seo, Seongjae Eom, Jong-min Lee
School of Nano Convergence Technology & Nano Convergence Technology Center, Hallym University, Chuncheon 24252, Republic of Korea
E-mail: jmlee@hallym.ac.kr

Jeong-Su Lee, Donghan Lee
Bright Quantum incorporated, Daejeon, 34134, Republic of Korea

Min Yong Jeon
Department of Physics, Chungnam National University, Yuseong-gu, Daejeon, 34134, Republic of Korea

# Contents

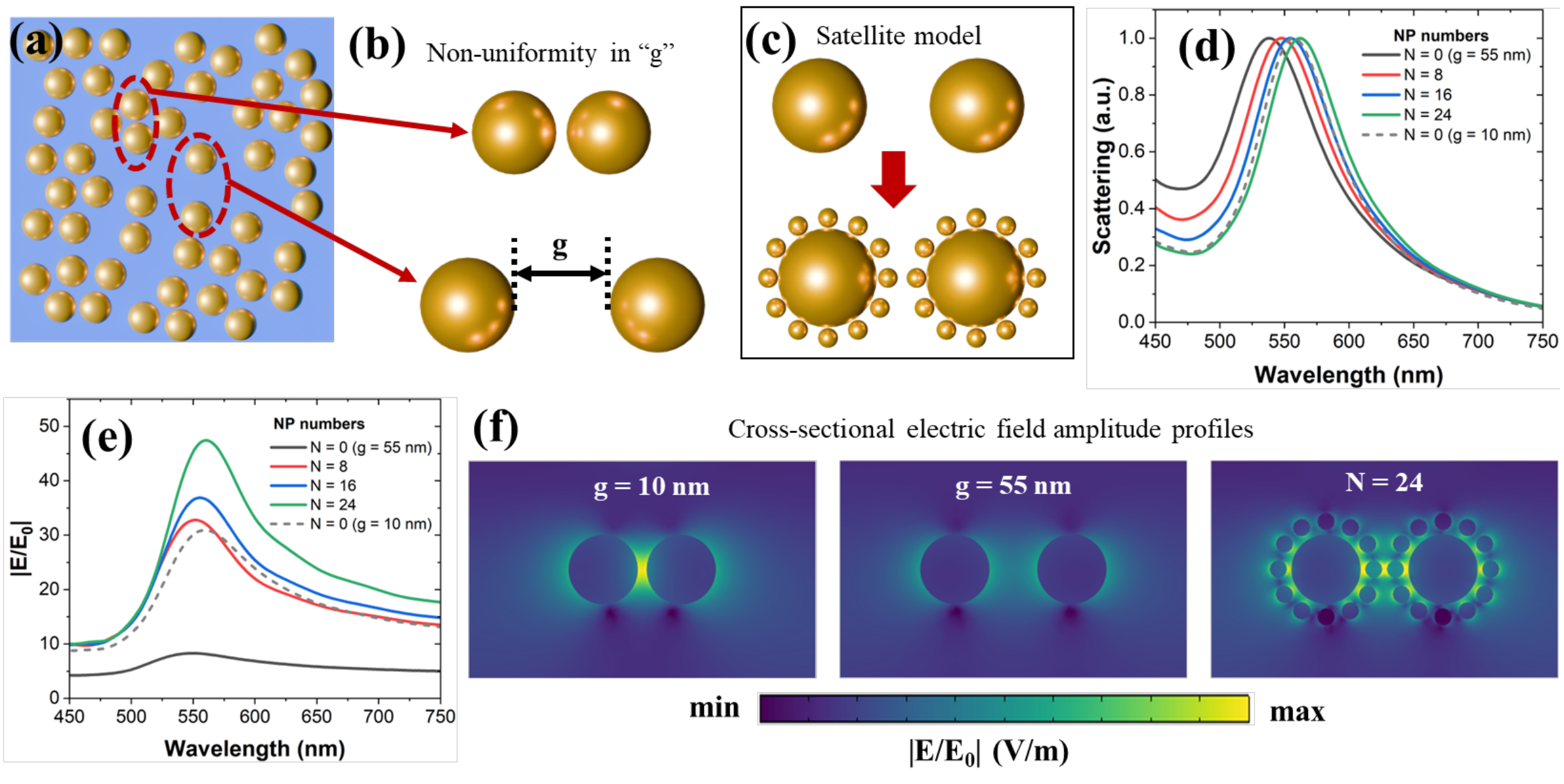

**Figure S1.** Many-body plasmonic cluster architecture with varied N number of satellite smaller nanoparticles (NPs). (a) Schematic illustration of self-assembly of plasmonic nanoparticles over a large area. (b) When observed closely in terms of dimer nanoparticles (dotted red oval), a non-uniformity (interparticle distance or gap – g) between nanoparticles displays a major issue which will deteriorate the plasmonic properties. (c) To solve this non-uniformity, satellite nanoparticles (smaller nanoparticles) can be introduced to fill the large gaps in a dimer. (d) Plasmonic scattering and (e) near-field enhancement |E/E0| results obtained from main manuscript figure 1a free space models. (f) Cross-sectional electric field amplitude profiles extracted at its resonant peak wavelength positions for dimer nanoparticles (N = 0) with g = 10 nm, 55 nm and with an introduction of satellite NPs filling N = 24 model.

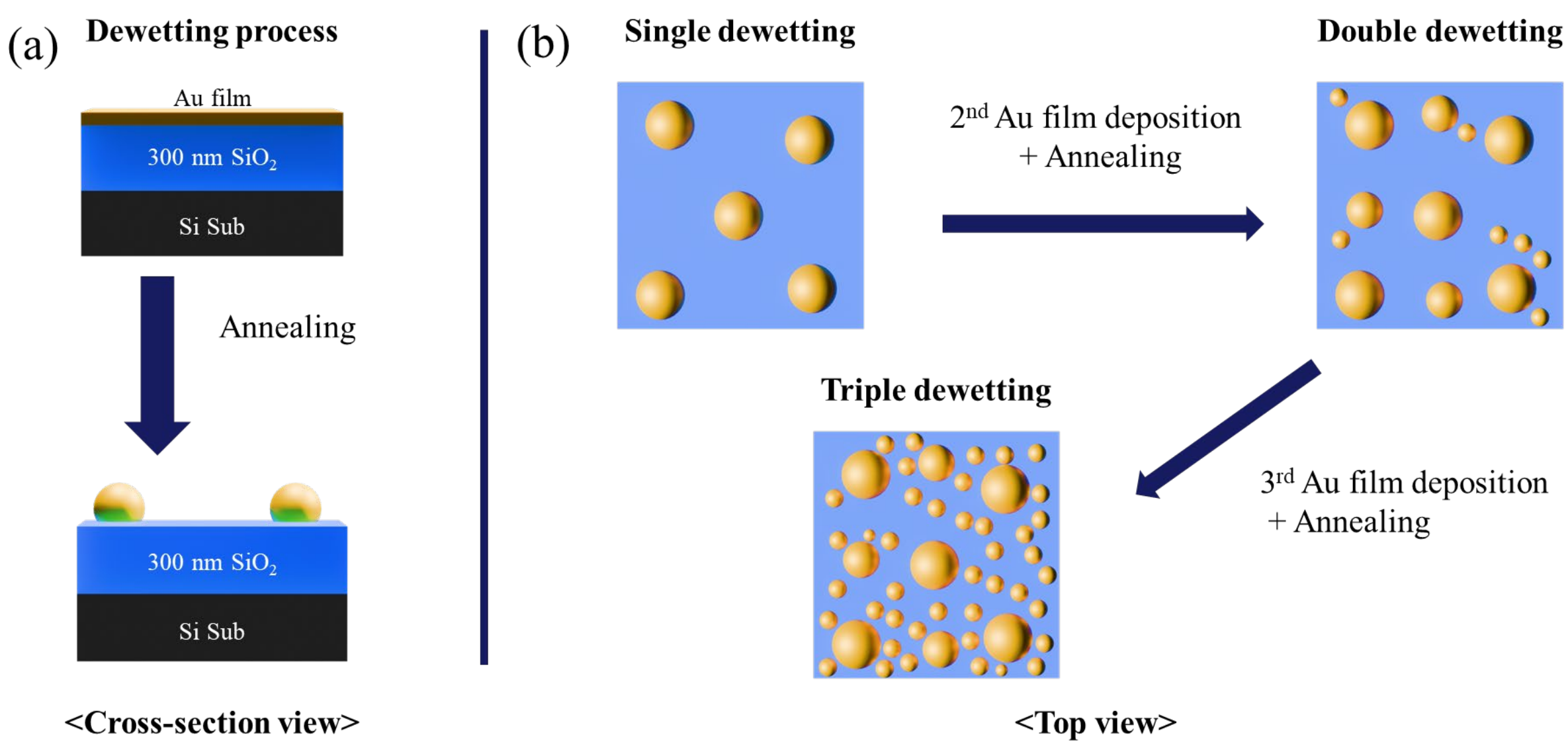

**Figure S2.** (a) Schematic illustration of our dewetting fabrication approach. (b) Single to triple dewetting fabrication process steps.

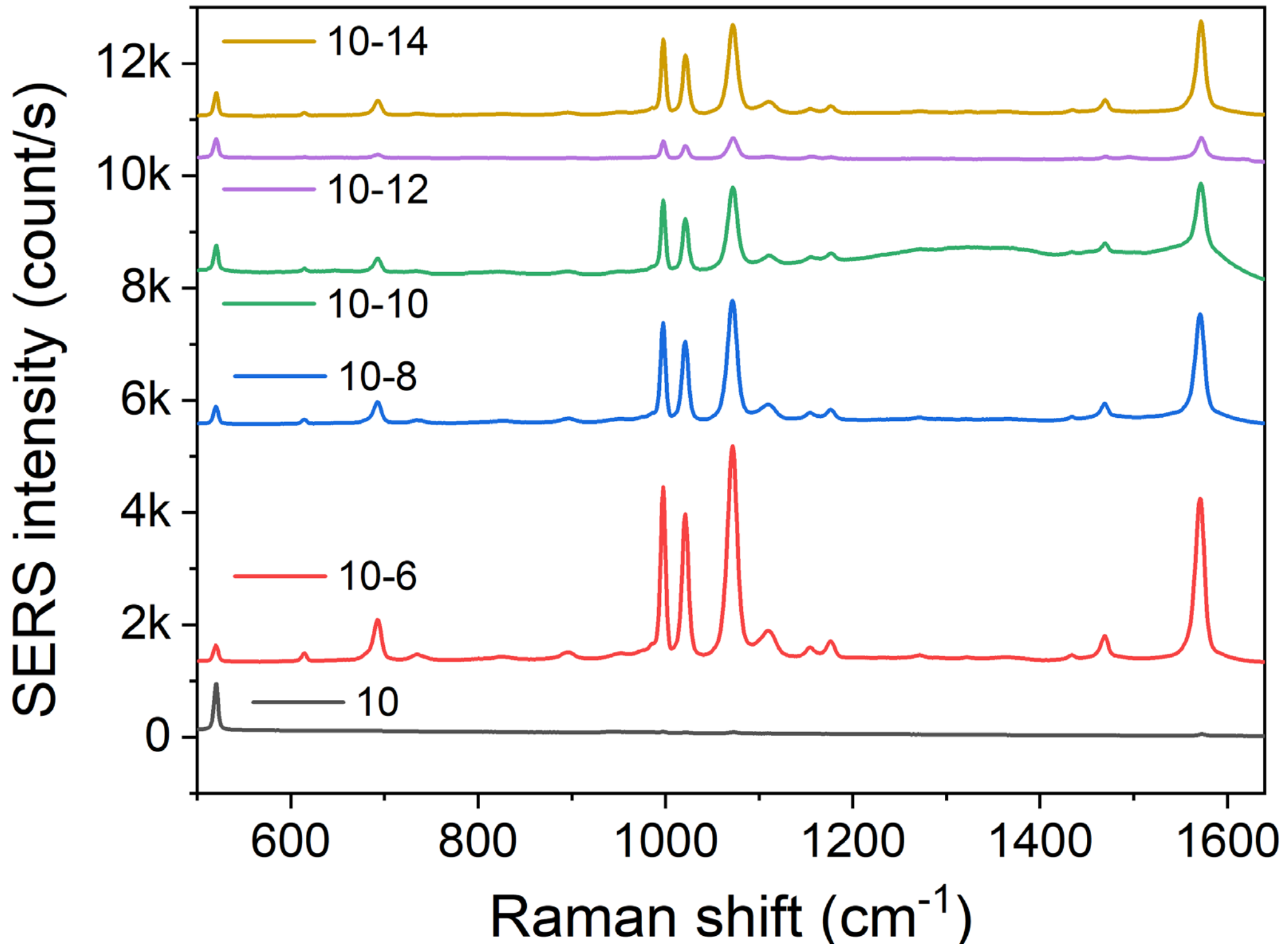

**Figure S3.** Raw SERS spectra obtained from double dewetted plasmonic nanostructures. Excitation wavelength 632 nm. The black solid line is taken from single dewetting condition with Au film thickness of 10 nm.

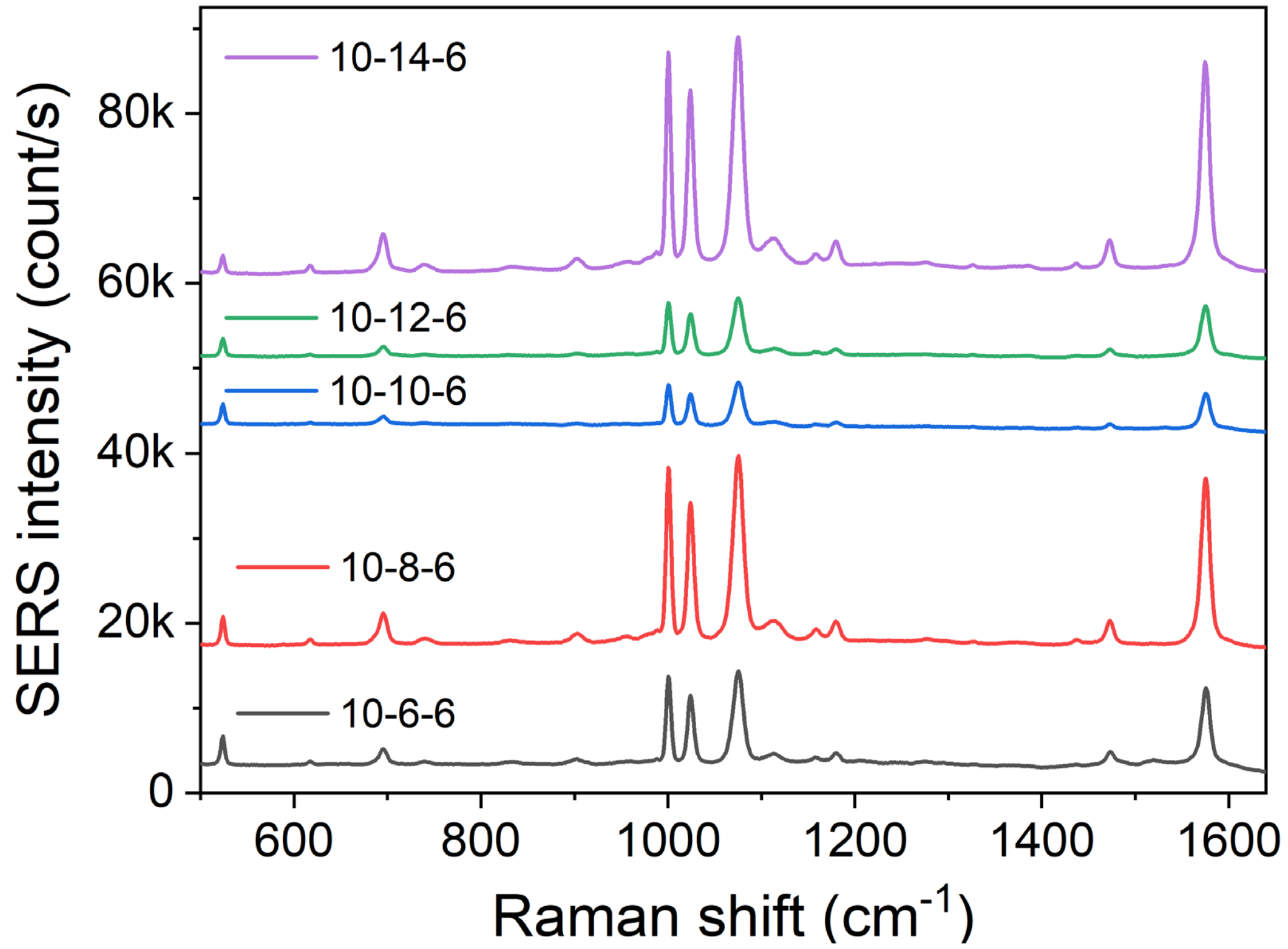

**Figure S4.** Raw SERS spectra obtained from triple dewetted plasmonic nanostructures. Excitation wavelength is 632 nm.

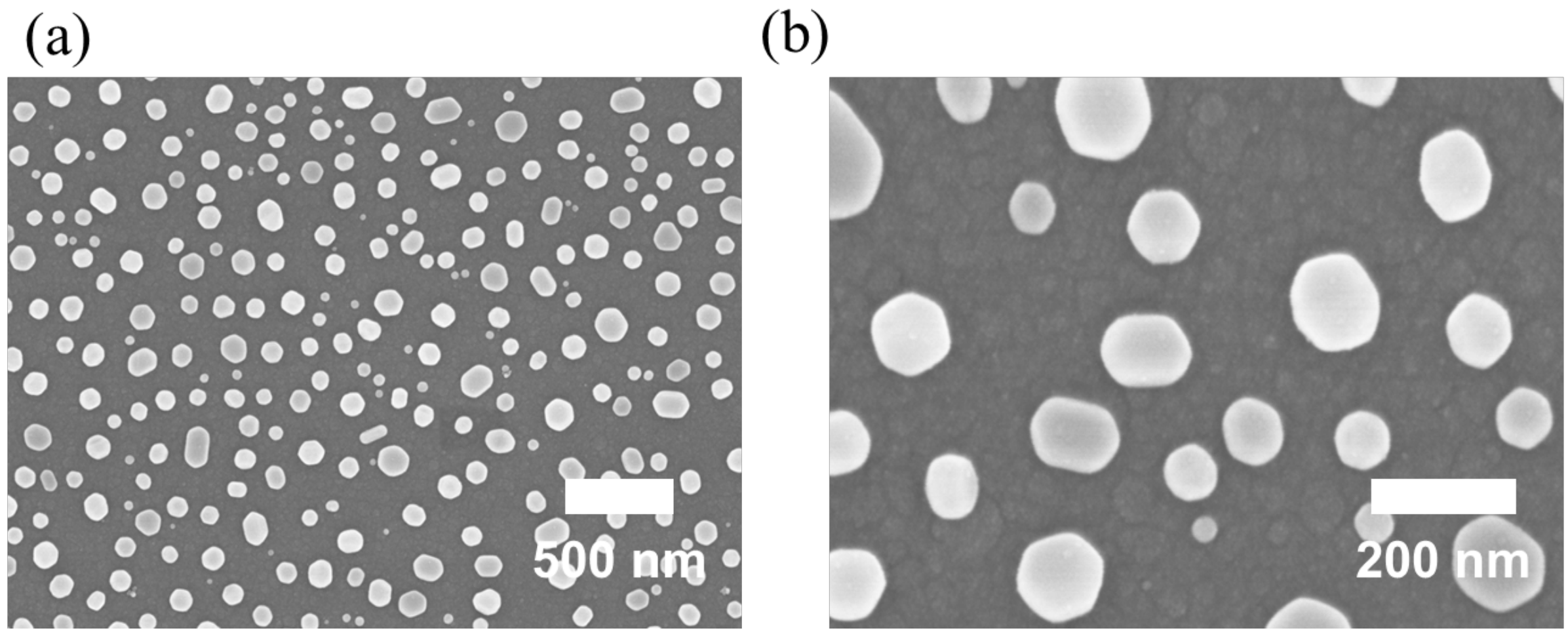

**Figure S5.** SEM images recorded at two different magnifications from single dewetted plasmonic nanostructures.

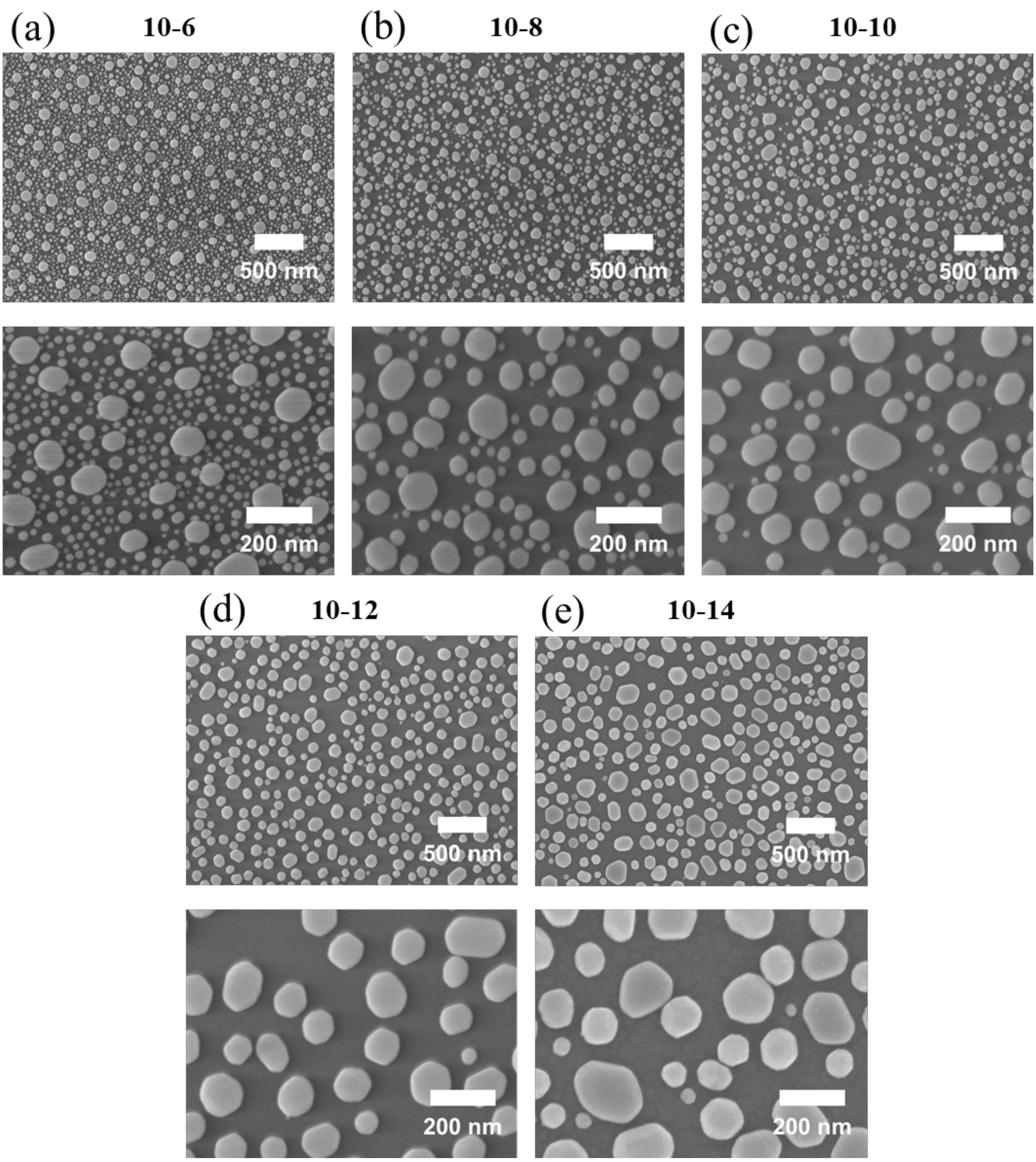

**Figure S6.** SEM images recorded at two different magnifications from double dewetted plasmonic nanostructures.

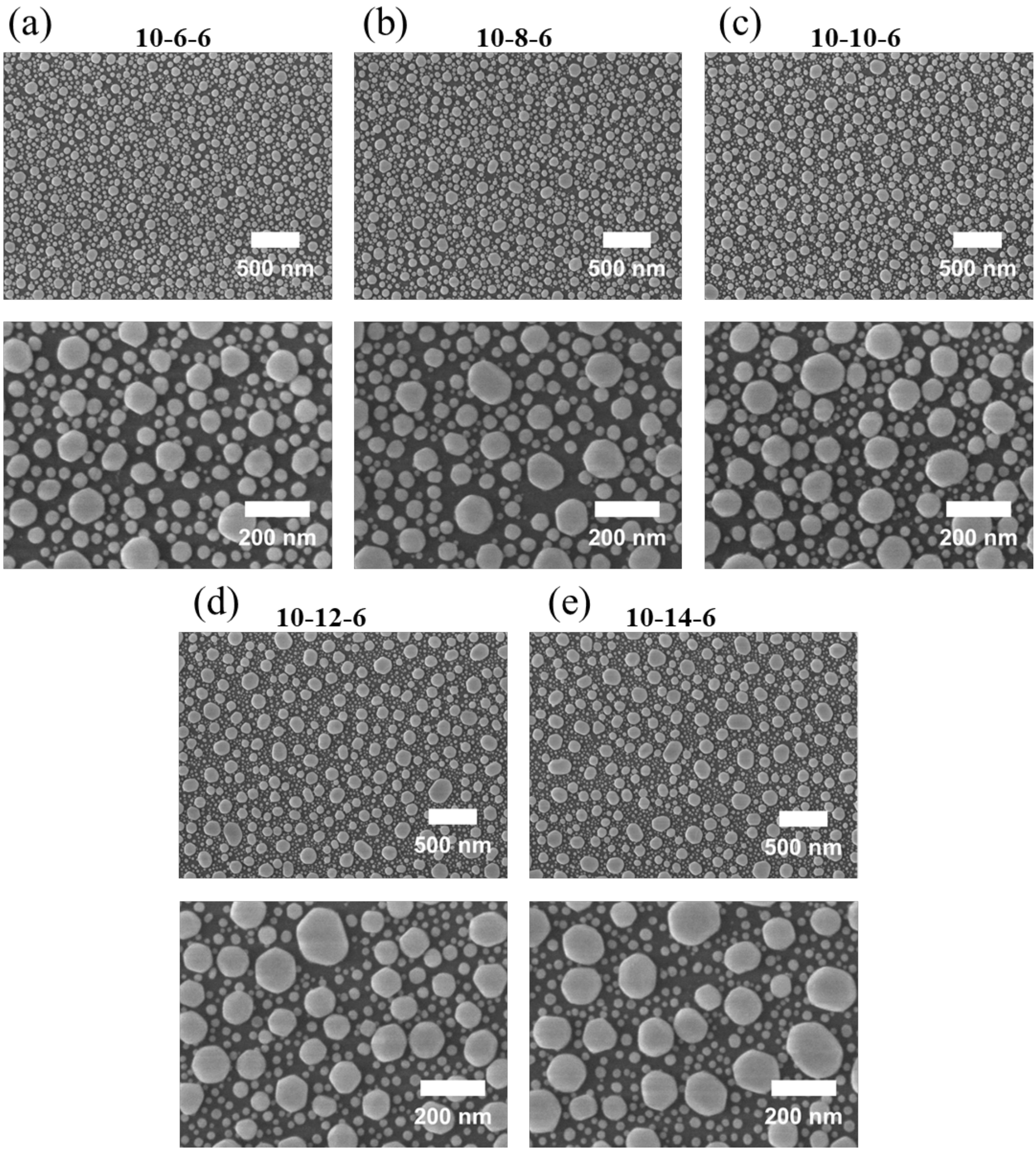

**Figure S7.** SEM images recorded at two different magnifications from triple dewetted plasmonic nanostructures.

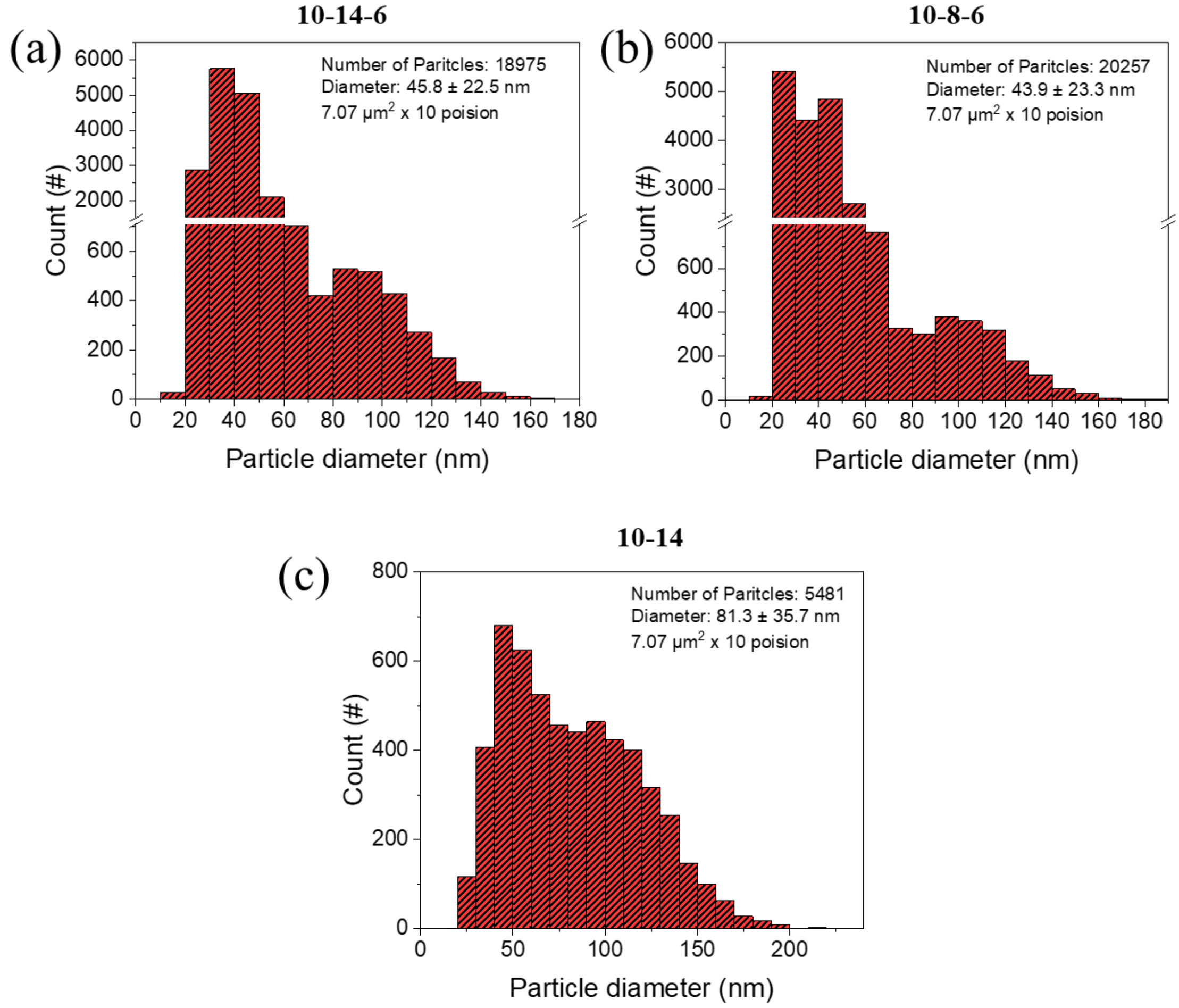

**Figure S8.** Nanoparticle (NP) diameter distribution from following dewetted plasmonic nanostructures: (a) 10-14-6, (b)10-8-6, and (c) 10-14.

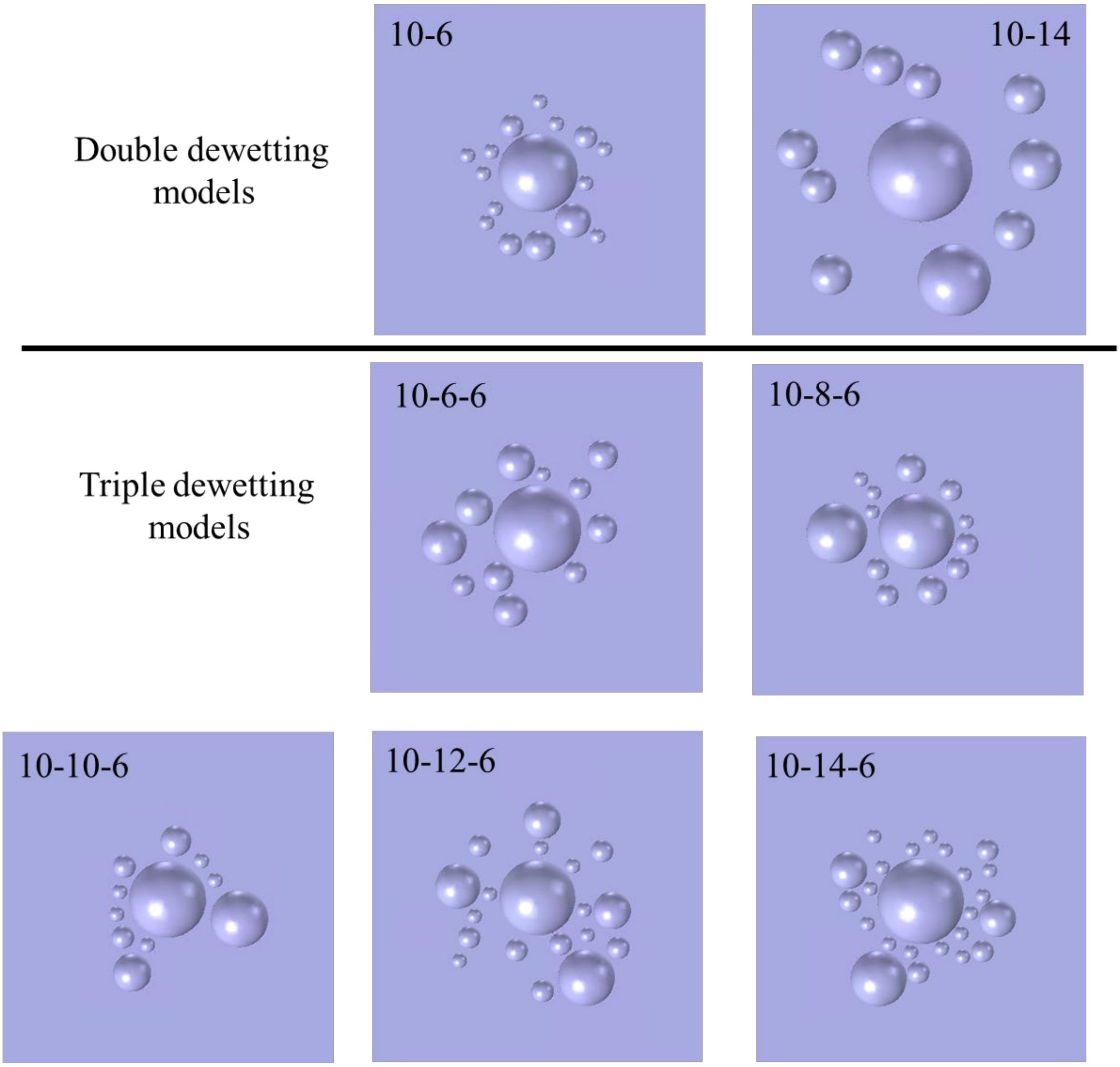

**Figure S9.** Simulated nanoparticle distribution models for double and triple dewetted plasmonic nanostructures as observed from SEM images.

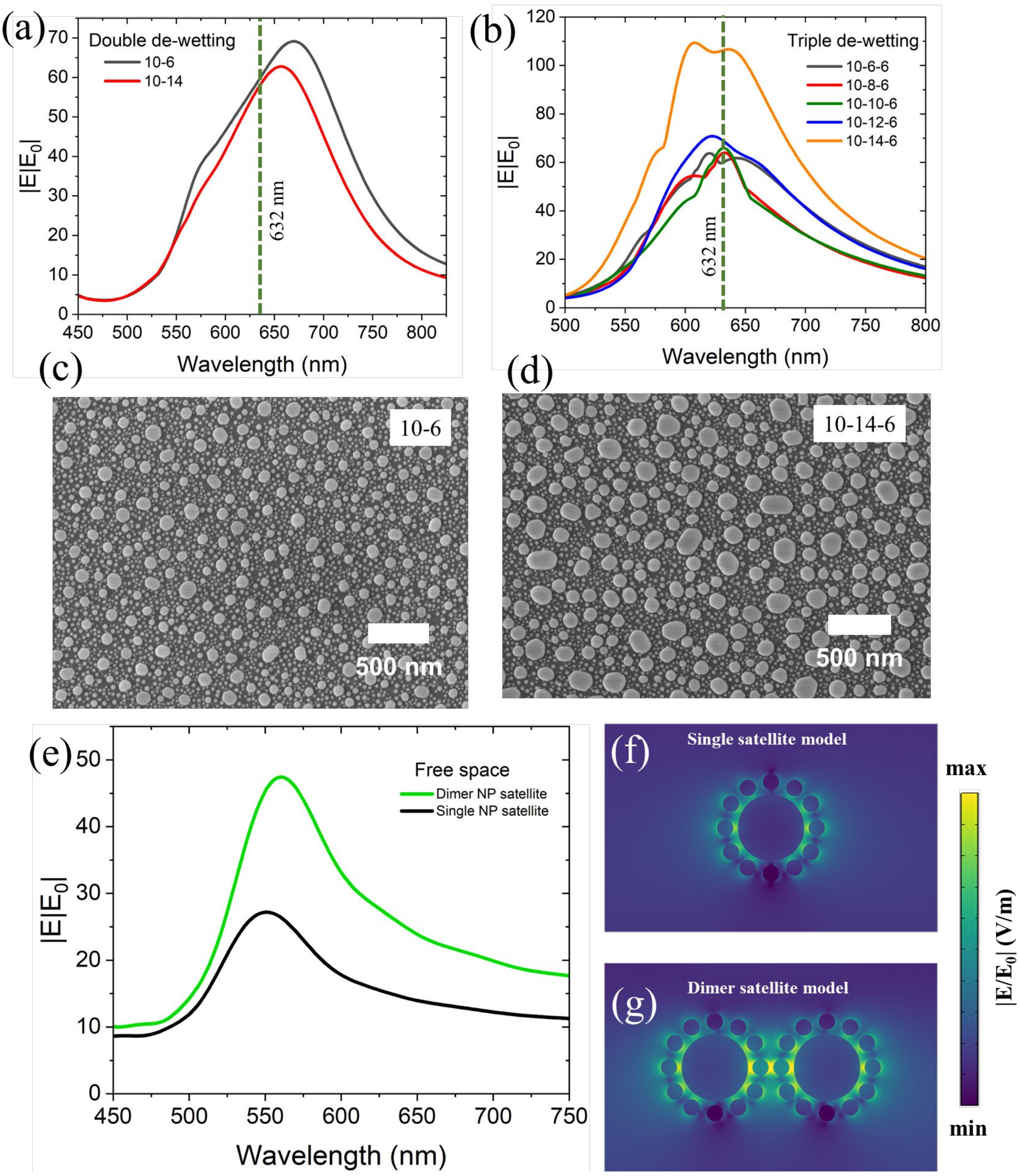

**Figure S10.** Simulated nearfield enhancement $|E/E_0|$ spectra for (a) double and (b) triple dewetted plasmonic nanostructures. SEM images displaying the differences between (c) 10-6 and (d) 10-14-6. Simulated nearfield enhancement $|E/E_0|$ (e) spectra and (f – g) cross-sectional XZ electric field amplitude profiles for satellite nanostructures surrounding single NP and dimer NP model.

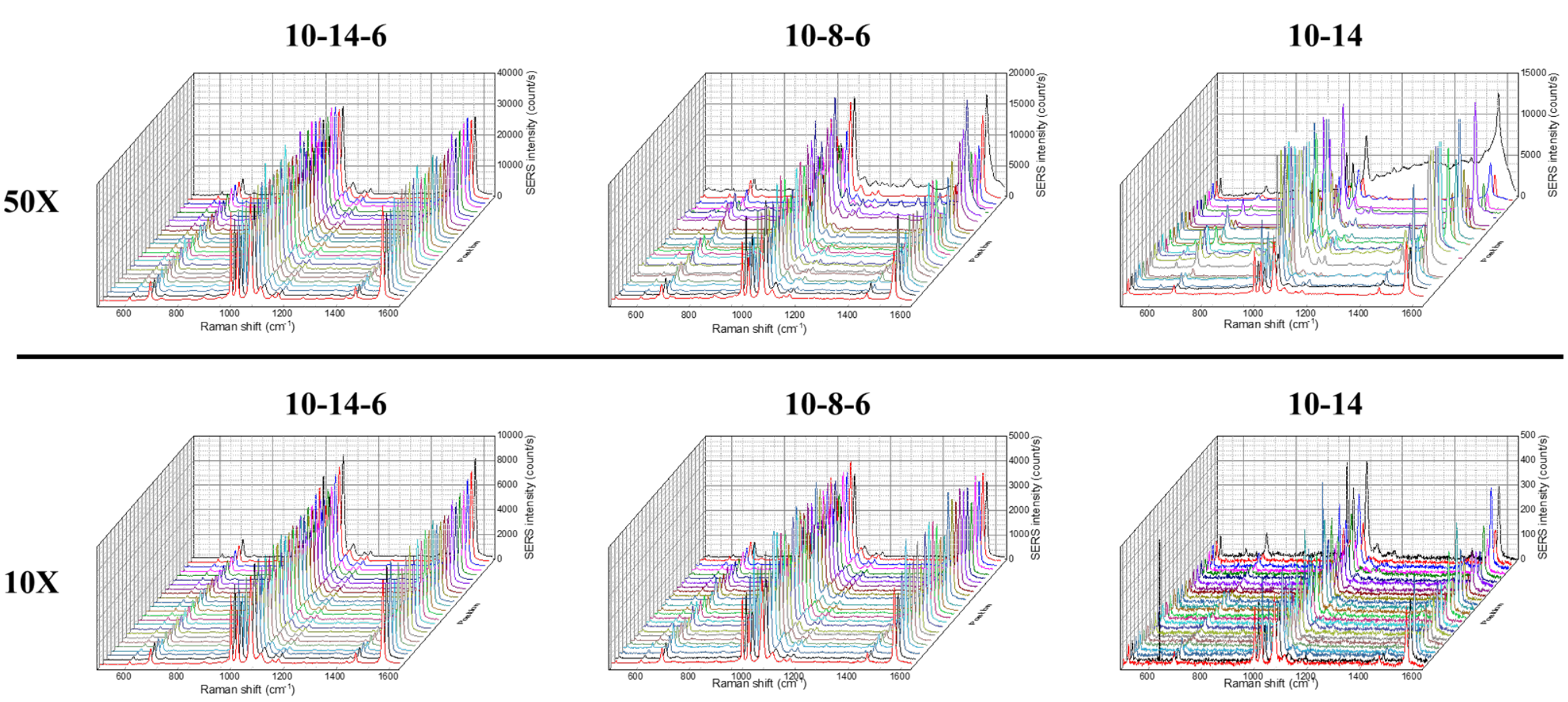

**Figure S11.** Raw SERS spectra taken with 10X and 50X objectives from 10-14-6, 10-8-6 and 10-14 dewetting substrates.

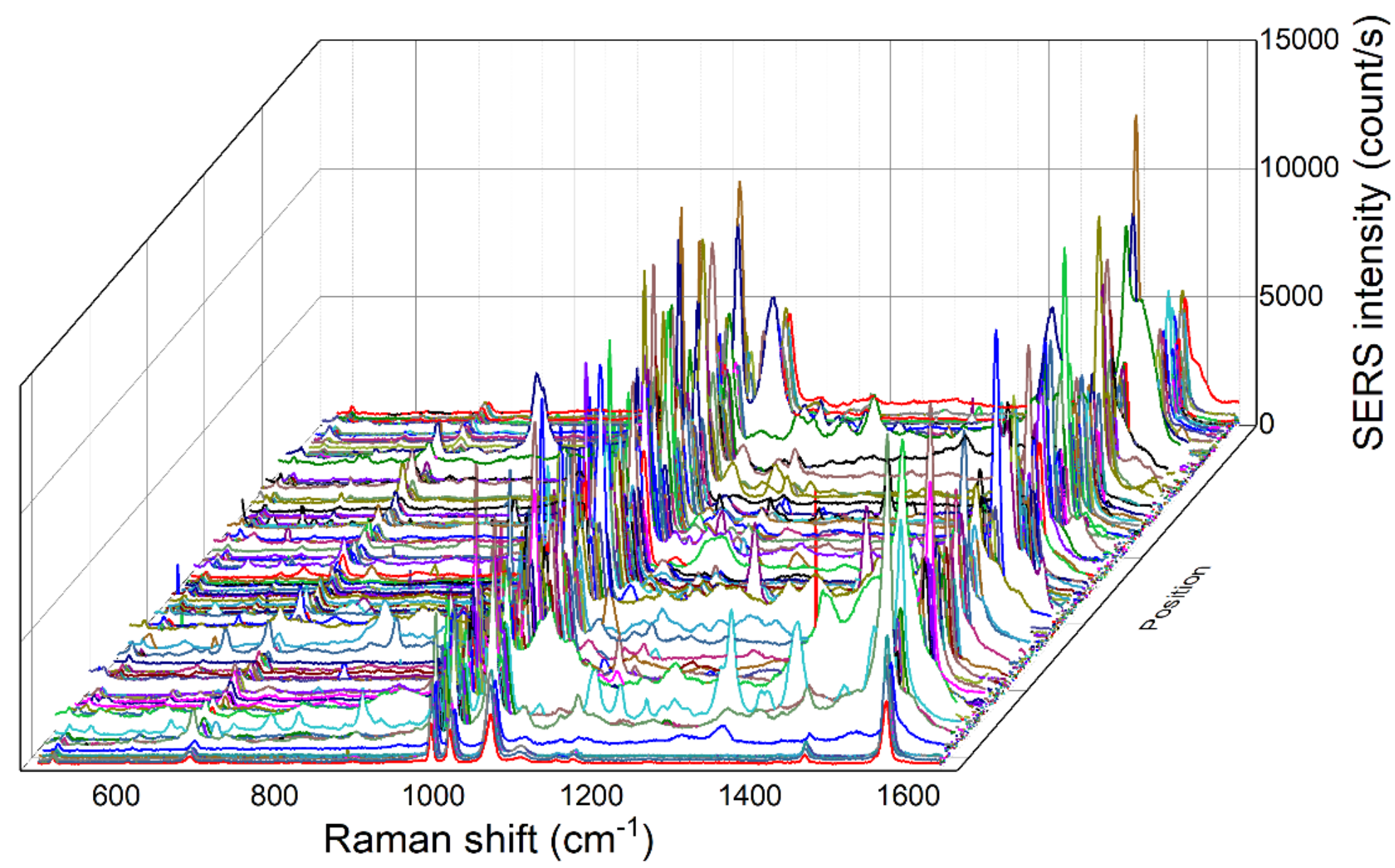

**Figure S12.** Raw SERS spectra used for high resolution mapping from 10-14 double dewetted plasmonic substrate.

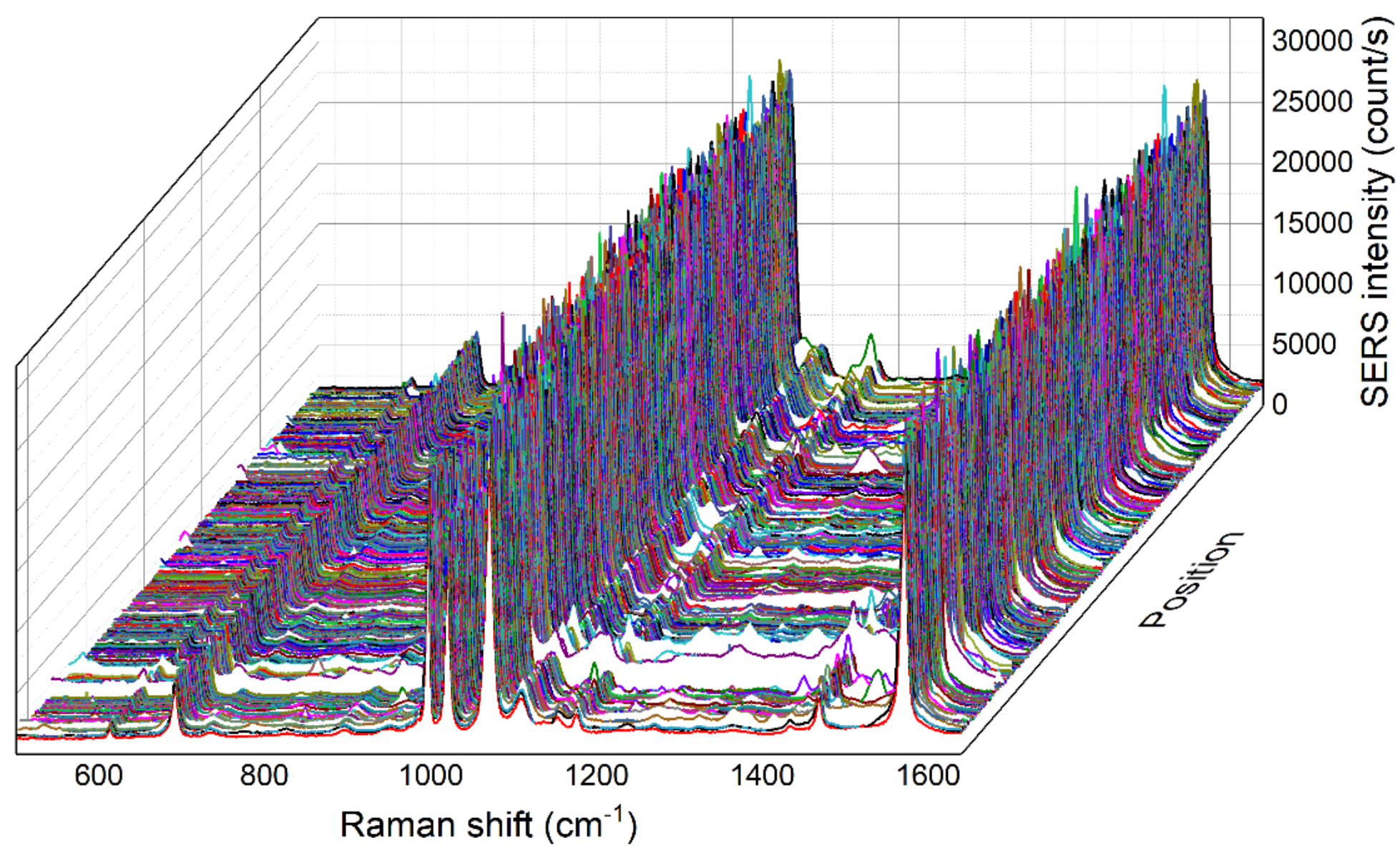

**Figure S13.** Raw SERS spectra used for high resolution mapping from 10-14-6 triple dewetted plasmonic substrate.

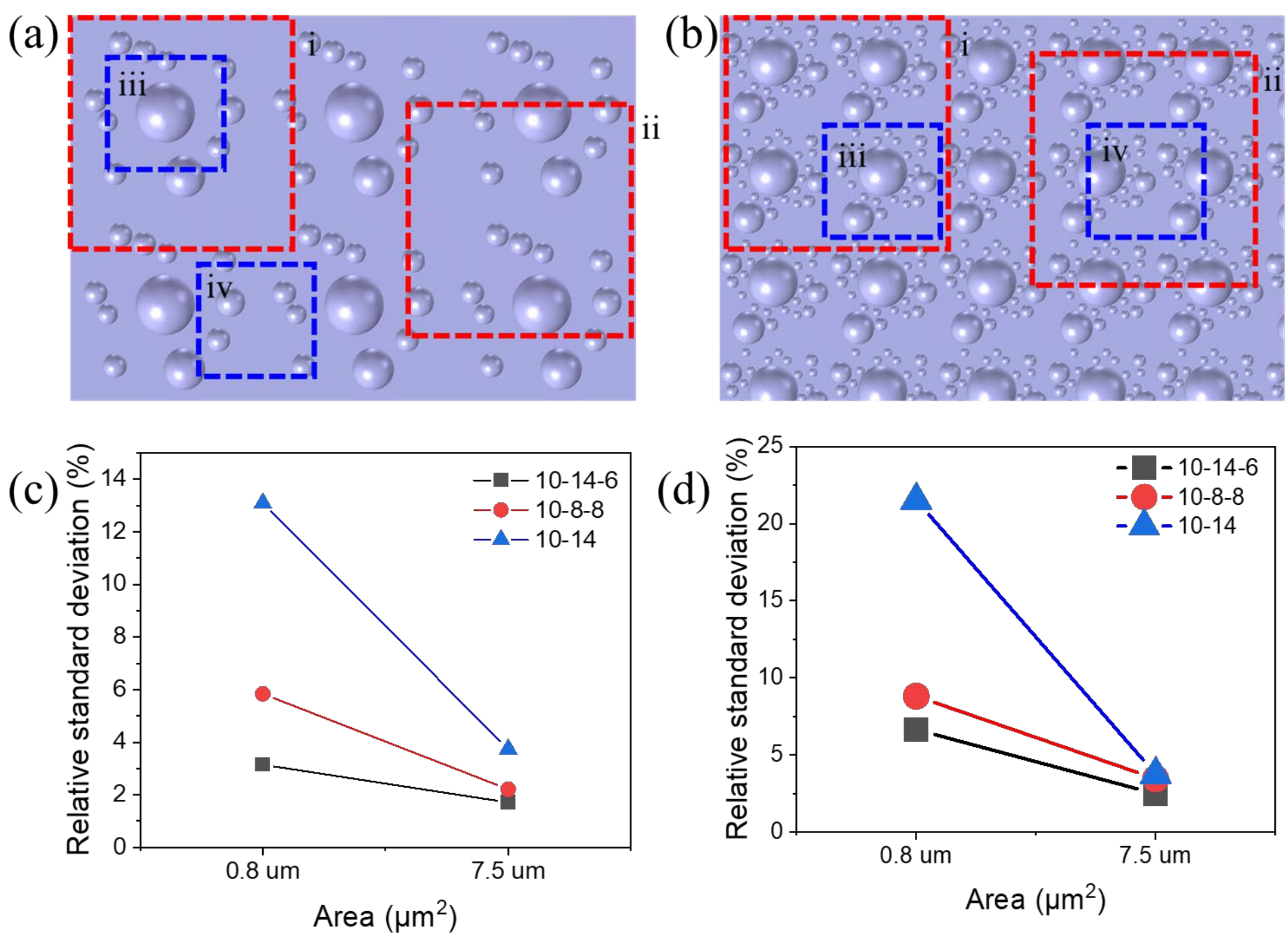

**Figure S14.** (a, b) Schematics illustrating the relationship between analysis area and nanoparticle uniformity on sample surfaces under dewetting conditions (a) 10-14 and (b) 10-14-6. Red boxes (i, ii) indicate larger analysis areas, whereas blue boxes (iii, iv) indicate narrower analysis areas. (c) Relative standard deviation (RSD) of average particle size across different analysis areas. (d) RSD of average interparticle distance across different analysis areas.

**Evaluation of SERS uniformity**: The correlation between analysis area and uniformity strongly depends on the characteristic size of the repeating nanoparticle structures. Both samples in figure S12 (a) and (b) exhibit repetitive nanoparticle arrangements; however, sample (a) has a larger unit cell size, causing smaller analysis areas (blue boxes iii and iv) to include partial or incomplete unit cells. As a result, these smaller analysis areas highlight local irregularities, leading to lower perceived uniformity. In contrast, sample (b), which features a significantly smaller unit cell, maintains high uniformity even in narrow analysis areas because these areas can fully capture complete repeating structures. Among the studied nanoparticle configurations, the 10-14-6 sample demonstrated the lowest variability and the highest enhancement in surface-enhanced Raman scattering (SERS) intensity. Specifically, within the 0.8 μm² analysis region, the relative standard deviation (RSD) of particle counts for the 10-14-6 sample was only 3.14%, significantly lower than the 5.84% and 13.1% observed in the 10-8-8 and 10-14 samples, respectively (Fig. S12(c)). Expanding the analysis region from 0.8 $\mu m^2$ to 7.5 $\mu m^2$ further reduced particle count deviations to 1.71% in the 10-14-6 sample. Interparticle gap distributions showed similar behavior. For the 10-14-6 sample, gap deviation decreased notably from 6.64% (0.8 $\mu m^2$) to 2.47% (7.5 $\mu m^2$), whereas the 10-8-8 and 10-14 samples improved from 8.8% to 3.4% and from 21.5% to 3.73%, respectively (Fig. S12(d)). Thus, the 10-14-6 sample's relatively small unit cell size ensures improved local uniformity, suggesting substantial potential to enhance the integration density of nanophotonic devices.

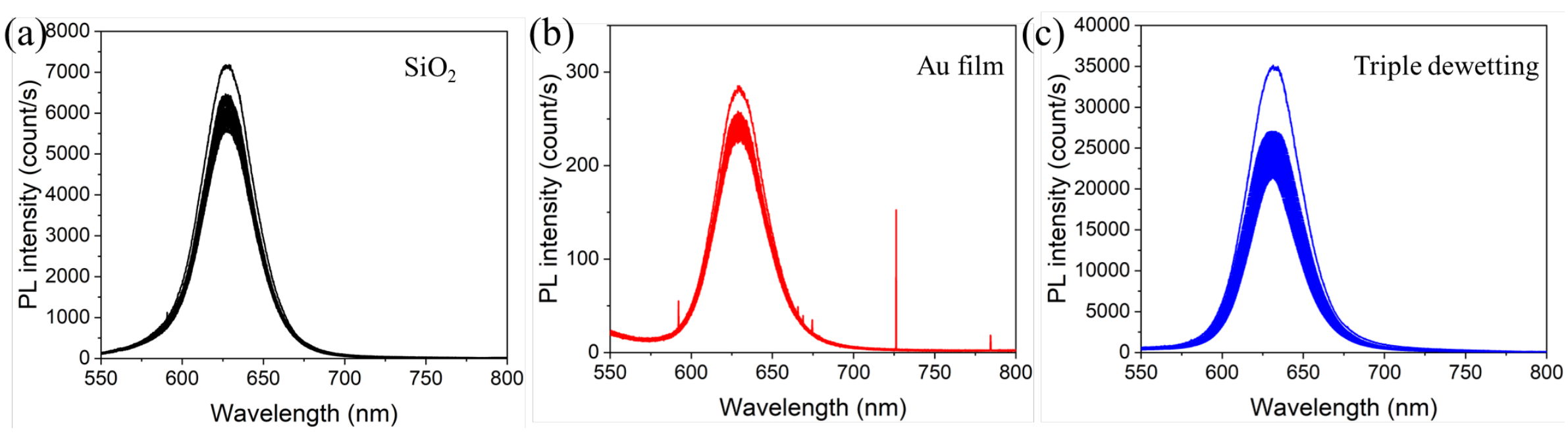

**Figure S15.** Photoluminescence characterization of InP quantum dots on different substrates: (a) on $SiO_2$, (b) Au film, and (c) triple dewetted plasmonic substrate. 25 measurements are taken at random positions for each substrate.

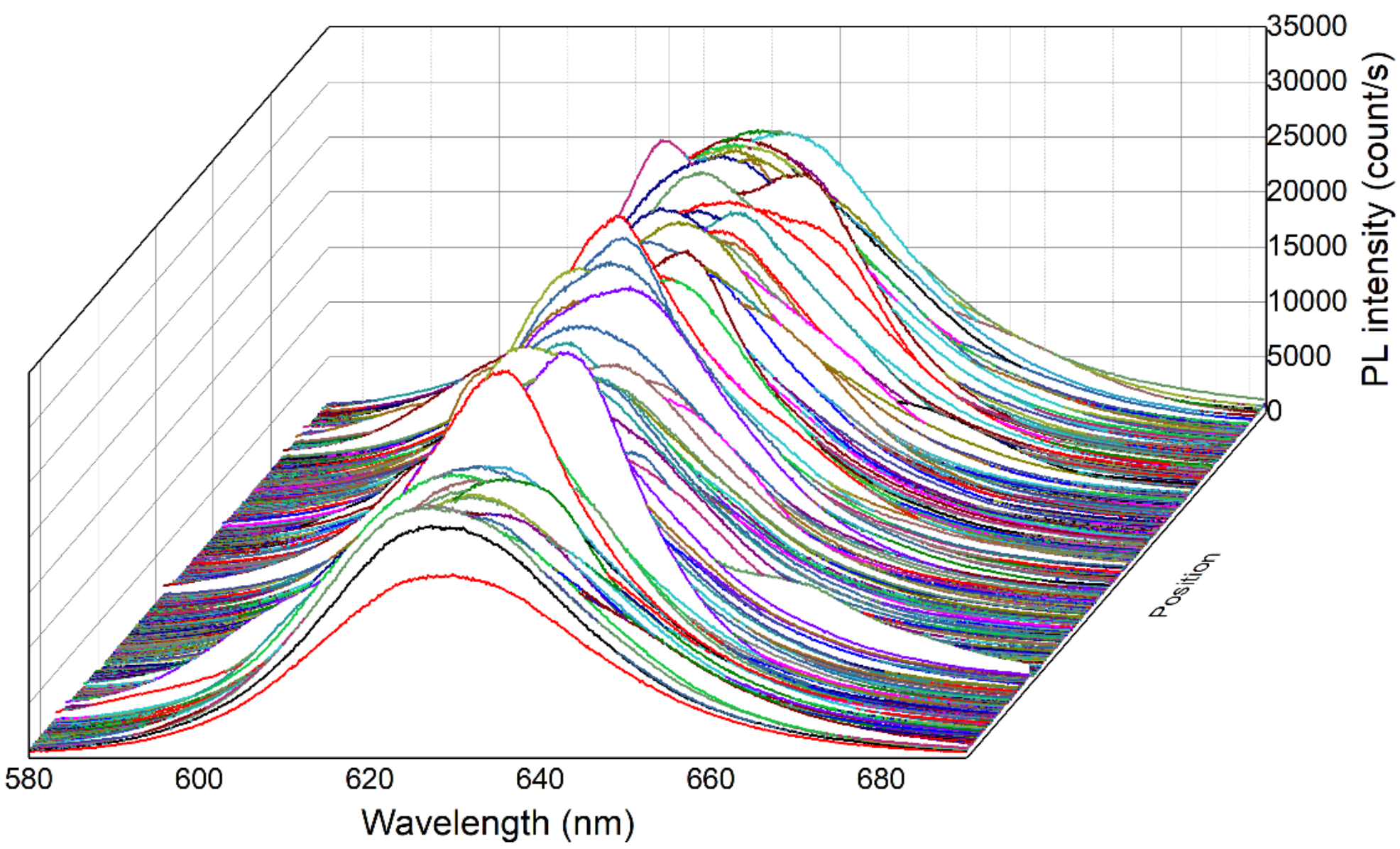

**Figure S16.** Raw photoluminescence spectra used for high resolution mapping from InP quantum dots on a 10-14-6 triple dewetted plasmonic substrate.